\documentclass[prd,aps,preprint,amsmath,nofootinbib,amssymb,eqsecnum,showkeys,tightenlines]{revtex4-1}

\usepackage{graphicx}
\usepackage{subcaption}
\usepackage{float}
\usepackage{amsmath}
\usepackage{amssymb}
\usepackage[utf8]{inputenc}
\usepackage{hyperref}
\usepackage{color}

\newcommand{\be}{\begin{equation}}
\newcommand{\ee}{\end{equation}}

\begin{document}

\title{The Boosted Higgs Jet Reconstruction via Graph Neural Network}

\author{Jun Guo$^1$}

\author{Jinmian Li$^2$}
\email{jmli@scu.edu.cn}

\author{Tianjun Li$^{3,4}$}

\author{Rao Zhang$^2$}

\affiliation{$^1$ Department of Physics, Jiangxi Normal University, Nanchang 330022, China}
\affiliation{$^2$ College of Physics, Sichuan University, Chengdu 610065, China}
\affiliation{$^3$ CAS Key Laboratory of Theoretical Physics, Institute of Theoretical Physics, Chinese Academy of Sciences, Beijing 100190, China}
\affiliation{$^4$ School of Physical Sciences, University of Chinese Academy of Sciences,
No.~19A Yuquan Road, Beijing 100049, China}

\begin{abstract}
By representing each collider event as a point cloud, we adopt the Graphic Convolutional Network (GCN) 
with focal loss to reconstruct the Higgs jet in it. 
This method provides higher Higgs tagging efficiency and better reconstruction accuracy than 
the traditional methods which use jet substructure information.  
The GCN, which is trained on events of the $H$+jets process, is capable of detecting a Higgs jet in events of several different processes, even though the performance degrades when there are boosted heavy particles other than the Higgs in the event. 
We also demonstrate the signal and background discrimination capacity of the GCN by applying it to the $t\bar{t}$ process. Taking the outputs of the network as new features to complement the traditional jet substructure variables, the $t\bar{t}$ events can be separated further from the $H$+jets events. 

\end{abstract}

\maketitle

\section{Introduction}

The precision measurements~\cite{Aad:2019mbh,Sirunyan:2018koj} of the Higgs boson couplings at the LHC are compatible with the Standard Model (SM) predictions. However, there are several evidences, such as dark matter observation and neutrino oscillation, indicating new physics Beyond the SM (BSM). In many of the SM extensions, the Higgs boson is a portal to BSM sector. Measuring the Higgs boson precisely at colliders is crucial for probing new physics.

The decay channel of $H\to b\bar{b}$ has not been probed with promising precision in spite of its large branching ratio, mainly due to the large QCD backgrounds. The jet substructure technique has been found to be efficient in tagging this channel when the Higgs boson is Lorentz-boosted. Various jet substructure tools and techniques~\cite{Abdesselam:2010pt, Altheimer:2012mn,Altheimer:2013yza,Adams:2015hiv,Larkoski:2017jix,Asquith:2018igt} have been developed since the first proposal in Ref.~\cite{Butterworth:2008iy}. 
In those methods, jets are reconstructed from sequential recombination algorithms, such as the anti-$k_T$ algorithm~\cite{Cacciari:2008gp}, the Cambridge/Aachen algorithm~\cite{Dokshitzer:1997in} and so on. 
Based on the distribution of jet constituents, several jet substructure variables are built to characterize the jet. 
This procedure involves a cone size parameter $R$ in jet clustering which should be adjusted for different searches, depending on the detector parameters and the target jet properties. 
As a result, hadrons with different origins may be clustered in a same jet if $R$ is too large, or those with the same origin may be mis-clustered into different jets if $R$ is too small. 
Moreover, in most jet substructure analyses, a large $R$ is chosen in order to fully capture the jet constituents during the jet clustering. It means that the jet is heavily contaminated by pileup events. 
A jet grooming method~\cite{Butterworth:2008iy,Krohn:2009th,Ellis:2009su} should be applied to mitigate the effects of pile-up events. 

Machine learning techniques have proven to be efficient in jet tagging and jet reconstruction (see Refs.~\cite{Guest:2018yhq,Albertsson:2018maf,Radovic:2018dip} for reviews). They can outperform the jet substructure method in tagging the Higgs jet~\cite{Datta:2019ndh,Lin:2018cin,Moreno:2019neq,Chakraborty:2019imr,Sirunyan:2020lcu}.  In Ref.~\cite{Li:2020grn}, we take the event information as a digital image and adopt the Mask R-CNN framework to detect/reconstruct the Higgs jet in the event image. The method achieves higher efficiency of Higgs jet detection and higher accuracy of Higgs momentum reconstruction than the traditional jet clustering and substructure tagging methods. 
In this work, we choose the Graph Neural Network (GNN) for Higgs jet reconstruction. 
Comparing to the imaged based methods, the GNN can be implemented in a more efficient way. 
The GNN methods classify final state hadrons with particular relations to form jets while the image based methods, such as the Mask R-CNN,  consist of two stages: image classification and region proposal.   
Despite its simpleness, the output of the GNN is more informative, {\it i.e.,} the output of GNN is particle wise. 
On the other hand, the Mask R-CNN only predicts continuous regions on the event image. All particles in a masked region are supposed to have the same origin. There are two drawbacks: (1) For charged particles, the tracker has better angular resolution than the electromagnetic/hadronic calorimeter. In experiments, a charged particle can be distinguished from other charged/neutral particles even when they are flying along a similar direction.
This feature can be easily encoded in GNN. However, the granulation in Mask R-CNN method is usually based on the cell size of calorimeters. All particles hitting the same calorimeter cell will be treated as a whole; 
(2) In the Mask R-CNN method, a calorimeter cell which contains only the particles from the pileup events but is surrounded by the Higgs constituents will also be assigned to the Higgs jet. The GNN does not have this tendency so that the pileup subtraction is less urgent. 

The GNNs have been applied to a growing variety of scenarios including social networks, knowledge graphs, recommender systems, 3D shape analysis, as well as the data analysis in high energy physics~\cite{Shlomi:2020gdn}. 
They have been used for pileup mitigation at hadron colliders~\cite{Martinez:2018fwc}, tracker reconstruction~\cite{Farrell:2018cjr,Ju:2020xty}, jet identification~\cite{Komiske:2018cqr,Qu:2019gqs,Moreno:2019bmu,Moreno:2019neq,Mikuni:2020wpr,Chakraborty:2020yfc,Bernreuther:2020vhm} and event classification~\cite{Choma:2018zbe,Abdughani:2018wrw}. In particular, studies in Refs.~\cite{Ju:2020xty,Ju:2020tbo} demonstrate that the GNN is capable of labelling constituents of a specific jet after supervised training. However, the effects from pileup events are not considered. They only need to label $\sim$10\% of total particles. 
In practice, with average number of $\langle n \rangle=50$ pileup interactions, the total number of particles for an event can reach $\mathcal{O}(10^4)$, whereas only $\lesssim \mathcal{O}(10^2)$ particles are labeled. This renders very low efficiency in training the GNN. 

In this work, we adopt the dynamic Graph Convolutional Network (GCN)~\cite{Wang:2018nkf} to reconstruct a boosted Higgs jet in any events, taking into account the pileup effects. 
The GCN is trained on the events of the $H+$jets process (with $p_T(H) >200$ GeV) overlaid with $\langle n \rangle=50$ pileup events. 
It is designed with focal loss~\cite{2017arXiv170802002L}, in order to train with the imbalanced dataset, {\it i.e.,} the number of unlabeled particles is around two orders of magnitude larger than that of the labeled ones.
We will demonstrate that the GCN is capable of detecting Higgs jet in events of several other processes efficiently, provided that there are no boosted heavy particles other than the Higgs in the event.  
Moreover, we will show that the GCN method has improved background discrimination power compared with the traditional jet substructure method.

The paper is organized as follows. In Sec.~\ref{sec:data}, we introduce how the event samples are generated and preprocessed. Some concepts related to the network are briefly introduced in Sec.~\ref{sec:model}.  
The performance of the network is discussed in Sec.~\ref{sec:result}. 
We provide conclusion and outlook in Sec.~\ref{sec:conclusion}.


 
\section{Event Simulation and Data Preparation}\label{sec:data}

Events in the analysis are simulated within MG5\_aMC@NLO~\cite{Alwall:2014hca} framework. Pythia8~\cite{Sjostrand:2007gs} is used to perform the parton shower, hadronization and hadrons decay. 
The events of $H$+jets with the Higgs boson decaying into a pair of $b$-quark are used for training and validating our GCN. 
The Higgs boson is required to have transverse momentum $p_T>200$ GeV. 
Moreover, there are multiple proton-proton collisions (referred to as pileup), during a bunch crossing period at the LHC. We adopt the A3 tune of Pythia8 with phenomenological parameters provided in Refs.~\cite{Skands:2014pea,ATLAS:2016puo} to simulate pileup events. 
In preparation of event samples, the final states of each event of a hard process are overlaid with final states of $\langle \mu \rangle =50$ (average number with poisson distribution) pileup events. This leads to dramatically increased difficulty in detecting the Higgs jet constituents with the GCN.  

\begin{figure}[htbp]
\centering
\includegraphics[width=6.5cm,height=6.5cm]{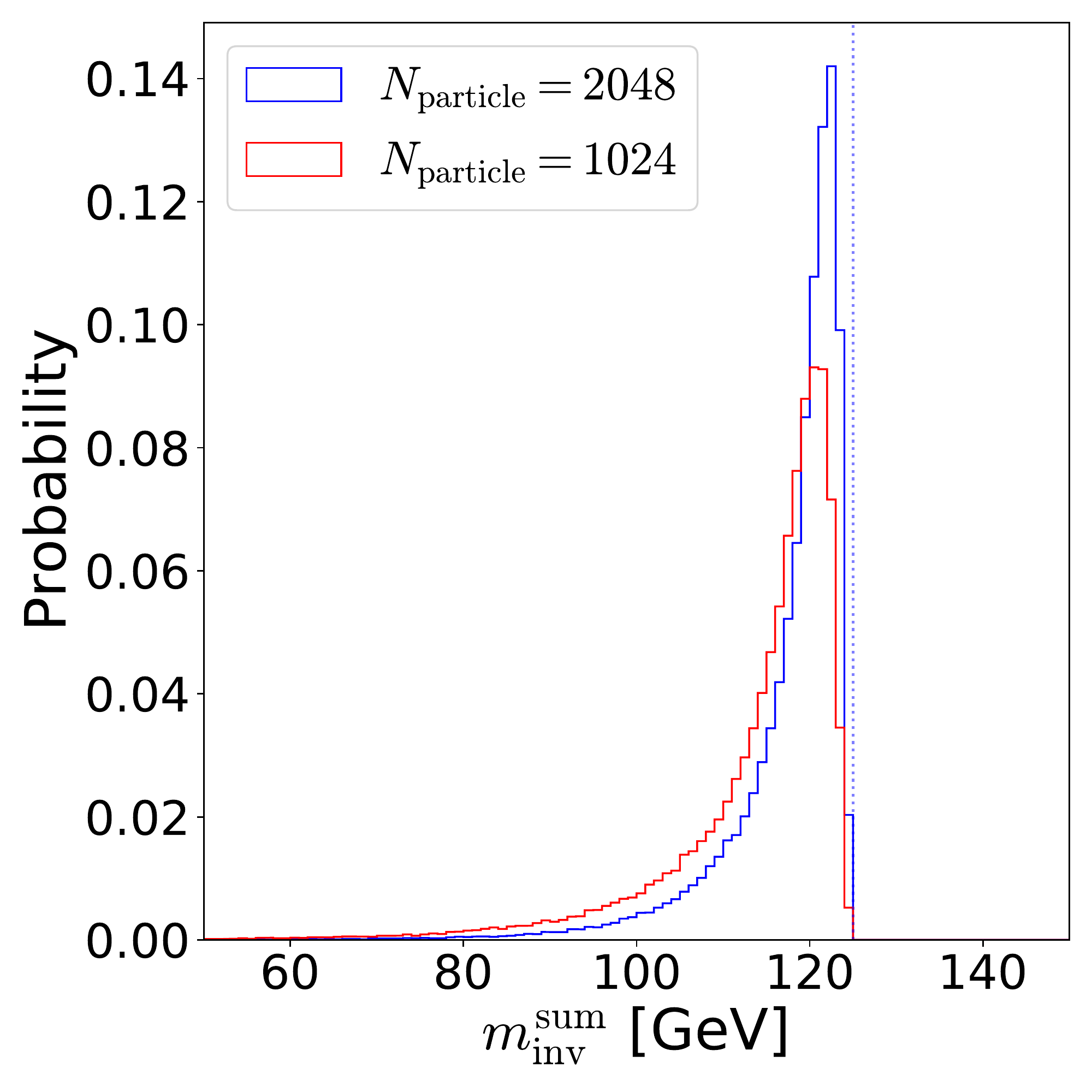}
\includegraphics[width=7.0cm,height=6.5cm]{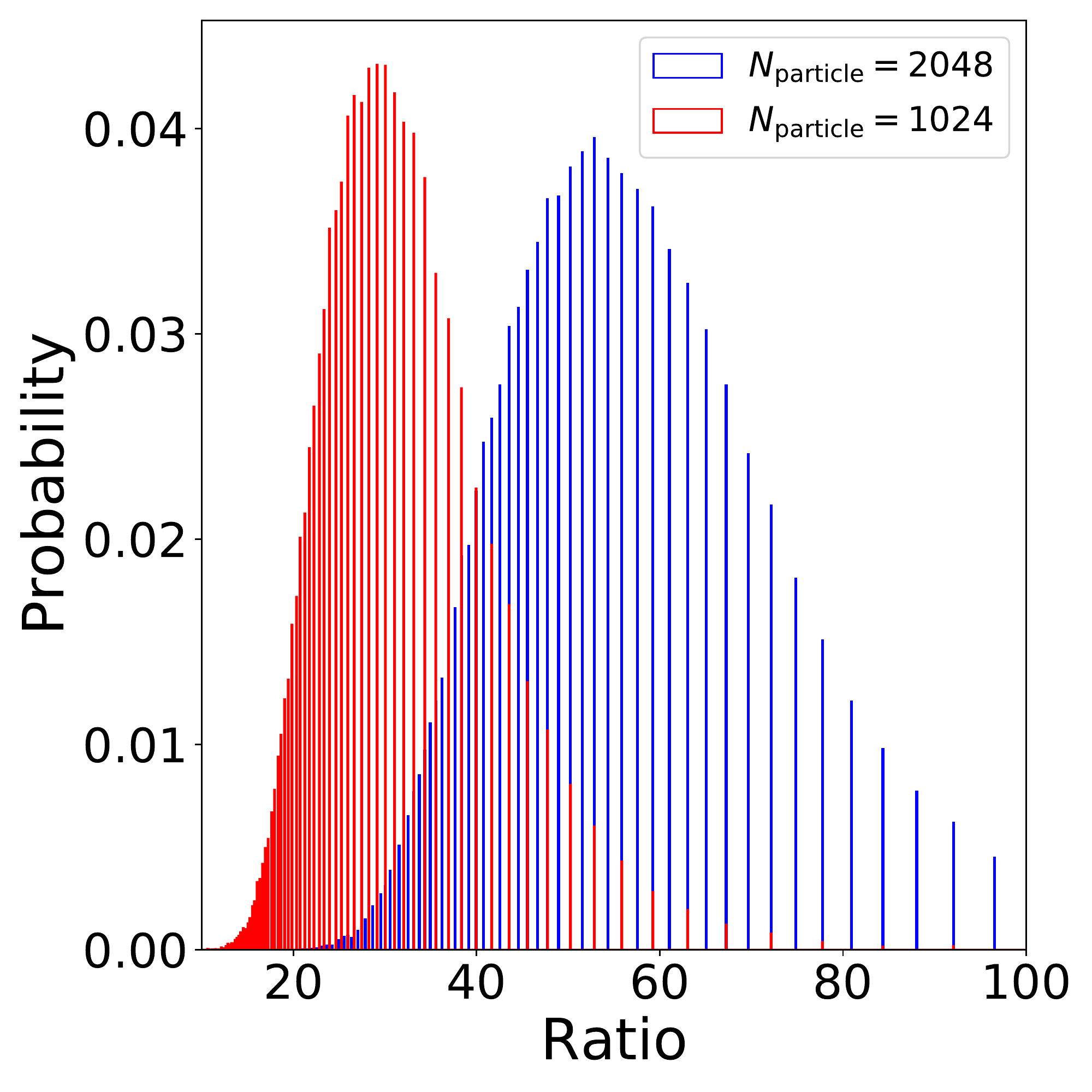}
\caption{\label{fig:samples} Left:  the invariant mass distributions for the vector sum of 
the selected Higgs constituents in two cases. 
Right: the ratio between the number of the unlabeled particles and that of the labeled ones. 
Events used here correspond to the $H$+jets process with $p_T(H)>200~\text{GeV}$  overlaid with pileup contaminations.  }
\end{figure}

With the contamination of pileup events, the total number of final state particles for each Higgs production event can reach $\mathcal{O}(10^4)$. 
However, we find that most of the Higgs constituents have relatively large transverse momenta. 
In order to reduce the complexity of the algorithm, we only consider the leading 1024 (or 2048) particles with the highest $p_T$ for each event. 
For the $H$+jets process with $p_T(H)>200$ GeV, the lowest $p_T$ of selected particle is $\sim 0.9$ GeV ($\sim 0.6$ GeV for selecting 2048 particles) for most of the events. So the selection can also help to alleviate the infrared effects of events.
In the left panel of Fig.~\ref{fig:samples}, we show the invariant mass distributions for the vector sum of the selected Higgs constituents momenta ($m^{\rm sum}_{\rm inv}$).  We find a certain distortion (deduction) in the invariant mass distribution due to our selection. 
The distortion is slightly less severe in the case with 2048 particles.
We will show later that such amount of distortion is acceptable in the sense that the GCN can perform better than 
the traditional jet substructure method. 
After the selection, the Higgs constituents still take only a small fraction of total particles. In the right panel of Fig.~\ref{fig:samples}, the distributions of the ratio between the number of the unlabeled particles (particles not belong to Higgs jet) and that of the labeled ones (Higgs constituents) are given. Because the typical number of Higgs constituents is 30-50, the ratio is $\sim$ 25 (50) in the case of selecting 1024 (2048) particles. 

The GCN takes the pseudo-rapidity ($\eta$), azimuthal angle ($\phi$), and transverse momentum ($p_T$) of all particles in an event as the input features. We have also tried to feed the particle type (lepton, photon or hadron) and particle electric charge to the GCN without finding much improvements~\footnote{The GCN is sensitive to the collinear splittings of input momenta. However, infrared effects of the hadronic level events are already alleviated in Monte Carlo simulation, since the very soft and collinear gluons in parton emission are not resolved in an infrared safe fragmentation framework, such as the string model~\cite{Andersson:1983ia} in Pythia8. For example, the MC simulations of the charged-particle multiplicities at the LHC are found to match the data well~\cite{Aad:2016mok}. (The charged particles are required to have $p_T \gtrsim 500$ MeV to improve the infrared stability.) And the number of charged particles in a jet is also used to distinguish quark and gluon jet at experiments~\cite{Cornelis:2014ima}.}.
Because the detector is a cylinder, the azimuthal angle ($\phi$, in the range of $[0,2 \pi)$)  is periodic. We find the GCN gains the best detection efficiency with the definition of $\phi$ such that the Higgs constituents do not distribute across the line of $\phi =0$. 
In the training sample, this condition is fulfilled by applying two pre-processing procedures: (1) The Higgs boson, whose azimuthal angle is outside the range of $[\pi/2, 3\pi/2]$, is shifted by $\pm \pi$ in the  $\phi$-coordinate to keep $\phi(H) \in [\pi/2, 3\pi/2]$; (2) The Higgs constituents, whose angular distances ($\Delta R = \sqrt{\Delta y^2 + \Delta \phi^2}$, $y$ is the rapidity) to the Higgs boson are larger than $\pi$/2, are not labeled. 
While during the testing, one can not obtain the angular position of the Higgs boson in advance. 
This condition can be fulfilled with the $\phi$ definition such that those high $p_T$ particles of an event are away from the line of the $\phi =0$. 
In addition, the training sample is purified by dropping events with $m^{\rm sum}_{\rm inv}$ smaller than 115 GeV, in order to not confuse the network with events that do not include any featured Higgs jets (All events are used in testing sample.).

Events of several processes different from the $H$+jets are also simulated to test the generality of the network: 1) two Higgs bosons plus three QCD jets; 2) one Higgs boson plus top quark pair; 3) a hypothetical Supersymmetric (SUSY) model process, $pp\to\tilde t^*_1\tilde t_1\to\bar t\tilde \chi_1^0 t \tilde \chi_2^0\to \bar t\tilde \chi_1^0 t H \tilde \chi_1^0$, with $m_{\tilde \chi_1^0}=100$ GeV, $m_{\tilde \chi_2^0}=800$ GeV and $m_{\tilde t_1}=1$ TeV.  
The Higgs bosons in the first two processes are forced to be boosted ($p_T>200$ GeV) and decayed into two $b$-quarks;
We note that each event of all those processes is overlaid with average number of 50 pileup events. 

\section{Graph Neural Network and Focal Loss}\label{sec:model}
The performance of the network in jet detection depends on both the data representation and the network architecture. 
In this section, we will give a brief introduction to the setups of our GCN and explain how focal loss works. 

\subsection{Event Representation and Graph Neural Network}

An event at the LHC corresponds to a collection of detected particles, with each particle assigned with a four-momentum ($\eta, \phi, p_T, E$) and an identity~\footnote{Only long-lived particles are detected. Electrons, muons and photons can be identified directly. Long-lived charged hadrons (pion, kaon and proton) can be distinguished through cherenkov effect, ionization energy loss and so on, while neutral hadrons are not distinguished from each other. However, the particle identities do not provide improved performance in our study. They will not be used as input to the GCN.}. 
The point cloud, which provides a flexible representation for collection of points in two or three dimensional space, can be naturally used for representing an event. 
Comparing to the image representation of collider events, which requires a uniform granulation, 
the different angular resolutions for different types of particles are easier to implement in the point cloud representation. 

Treating the point cloud as a graph, GCN is an efficient network to analyze it. 
A graph consists of nodes and edges. For collider events, nodes correspond to particles, and edges are the connections between particles.
Our GCN uses the connections of the $K$-nearest-neighbours (KNN) for each node in the feature space, with the squared distance defined as the sum of squared differences in all coordinates.  In the first edge-convolution (EdgeConv) block, it is
\be
D(x_i, x_j) = \sqrt{ (\eta_i - \eta_j)^2 + (\phi_i - \phi_j)^2}~.
\ee
As for the second and third EdgeConv blocks, the feature vectors learned by previous block are viewed as new coordinates of the nodes in the latent space. 
The neighbors of each node will be dynamically updated with EdgeConv operations.

In the edge-convolution blocks, the edge feature is calculated by~\cite{Wang:2018nkf,Qu:2019gqs}
\be
e_{ijm} = \mathrm{ReLU}(\boldsymbol{\theta}_m \cdot ((\mathbf{x}_{j}-\mathbf{x}_i) \oplus  \mathbf{x}_i)),
\ee
where index $i$ runs over all particles in an event (which is 1024 in our case), index $j$ runs over the $K$ neighbours of the particle $i$, and  index $m$ corresponds to the number of convolutional filter. 
The direct sum indicates that the feature vector of $i$-th particle ${\bf x}_i$ (with dimension $C$) being concatenated with the vector difference between the feature vectors of particle $i$ and its $j$th neighbour. $\boldsymbol{\theta}_m$ is a $2C$-dimensional vector which should be adjusted during the training. 
The inner product between $\boldsymbol{\theta}_m$ and the concatenated feature vector is delivered to the Rectified Linear Unit (ReLU) activation function to given the final expression. 
An aggregation function is applied after obtaining the edge features:
\begin{align}
x'_{im} = \max_{j} \ e_{ijm}. 
\end{align}
The new feature $x'_{im} $ will be passed to the next edge-convolution block. 

\begin{figure}[htbp]
  \begin{center}
       \includegraphics[width=1.0\textwidth]{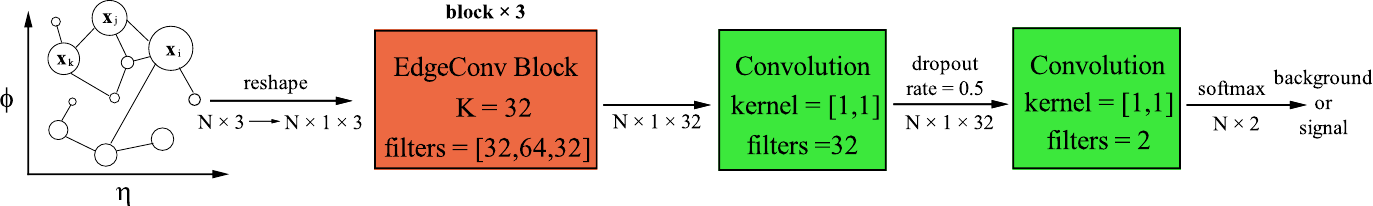}
    \caption{Architecture of our GCN, the edge-convolution block is the same as the one proposed in Ref.~\cite{Qu:2019gqs}. 
 \label{fig:network}}
  \end{center}
\end{figure}

The architecture of our GCN is illustrated in Fig.~\ref{fig:network}. The input coordinates are $\eta$ and $\phi$, and
the input features are $\eta$, $\phi$ and $p_T$. It contains three EdgeConv blocks, with a universal $K=32$. 
Each EdgeConv block is implemented as three multilayer perceptrons, which is the same as the one proposed in Ref.~\cite{Qu:2019gqs}. The number of channels (filters) of these perceptrons are  32, 64 and 32, receptively. 
Two convolutional layers with kernel size $[1, 1]$ are used for classification task. The first convolution layer contains 32 kernels, and the second contains 2 kernels. To prevent overfitting, a dropout layer with drop rate of 0.25 is inserted.
The final classification scores for input particles are assigned by the softmax function.

Note that usually a spatial transformer network (STN)~\cite{DBLP:journals/corr/JaderbergSZK15} is inserted in front of edge-convolution block to make the GCN invariant under translation, scale and rotation of input data. However, in our case, the coordinates $\eta$ and $\phi$ are not fully interchangeable, {\it i.e.} they have different distributions for given processes. We find including the STN module in the GCN does not improve the Higgs jet detection efficiency, as long as the training sample is sufficient. So the STN module is not adopted in our GCN. 
Moreover, by taking the $\eta$, $\phi$ and $p_T$ as input features for each particle (node), our GCN is sensitive to the position and the orientation of the graph, and each event can be associated with a graph without ambiguity.

\subsection{Focal Loss}
Focal loss was firstly introduced in Ref.~\cite{lin2018focal}, which is especially useful in object detection task. 
It is used for addressing the large class (signal and background) imbalance issue, and makes the model focus on signals which have much smaller number than backgrounds. 
It is a modification of cross entropy, given as
\be
FL(p, y) = -\alpha y(1-p)^\gamma \log(p) - (1-\alpha)(1-y)p^\gamma \log(1-p)~,
\ee
where $p$ is the probability of being positive class, and $y$ is the label which will be given beforehand in the supervised learning. Compared with cross entropy, the hyper parameter $\alpha$ is used to keep a balance between positive and negative samples, and $(1-p)^\gamma$ and $p^\gamma$ terms suppress the loss contribution of well classified samples so that loss function is dominated by the difficult ones.
For our case, given the ratio between the number of unlabeled particles and number of labeled ones around 30-50, 
we find setting $\alpha=0.65\ (0.75)$, $\gamma=1$ for $N_{\rm particle}=1024 \ (2048)$ samples works well.

\section{Result}\label{sec:result}
Our GCN model is implemented with TensorFlow~\cite{10.5555/3026877.3026899}. The model is trained on one million $H+$jets events with $p_T(H)>200$ GeV. The learning rate is set to $10^{-3}$. In this section, we present some results obtained by the well trained GCN.

\subsection{Number of input particles}
As have been discussed in Sec.~\ref{sec:data}, we consider input datasets with number of particles $N_{\text{particle}}=1024$ and $N_{\text{particle}}=2048$ for each event, respectively. 
We note that the GCN does not require a fixed size of input. So the GCN which is trained on either $N_{\text{particle}}=1024$ or $N_{\text{particle}}=2048$ event sample can be tested on both samples. There will be totally four different applications. We provide the recall (number of correct signal/number of predicted signal), accuracy (number of correct predictions/total number of predictions) and precision (number of correct signal/number of actual signal) for those cases in Tab.~\ref{tab:cases}. 
It is not surprised to find out that the recalls are the highest for the cases where the training and testing samples have the same input size. 
The accuracies are always high because most of the particles are belong to the background and they are easy to classify. 
Even though the event sample with larger $N_{\text{particle}}$ provides more complete list of Higgs constituents, the fraction of signal is significantly smaller (as shown in Fig.~\ref{fig:samples}). As a result, the precision is higher (the detected signal is purer) when the GCN is tested on the event sample with $N_{\text{particle}}=1024$. 
In the following, we will only consider the first case.

\begin{table}[htb]
  \begin{center} \renewcommand\arraystretch{1.3}
  \begin{tabular}{|c|c|c|c|c|c|} \hline
   & $N_{\rm train}$  & $N_{\rm test}$ & Recall & Accuracy & Precision \\
   \hline
   I &   1024 & 1024 & 0.699 & 0.979 & 0.724  \\[1mm]
   II &   1024 & 2048 & 0.635 & 0.984 & 0.604 \\[1mm]
   III &   2048 & 1024 & 0.645 & 0.976 & 0.673 \\[1mm]
   IV &   2048 & 2048 & 0.709 & 0.985 & 0.599 \\[1mm] \hline
  \end{tabular} \renewcommand\arraystretch{1.0}
  \caption{The recall, accuracy and precision of the GCN that is trained on $N_{\text{particle}}=N_{\text{train}}$ event sample, and tested on $N_{\text{particle}}=N_{\text{test}}$ event sample.  \label{tab:cases}}
  \end{center}
\end{table}

\subsection{Reconstructed kinematic variables}
\begin{figure}[htb]
  \begin{center}
       \includegraphics[width=0.42\textwidth]{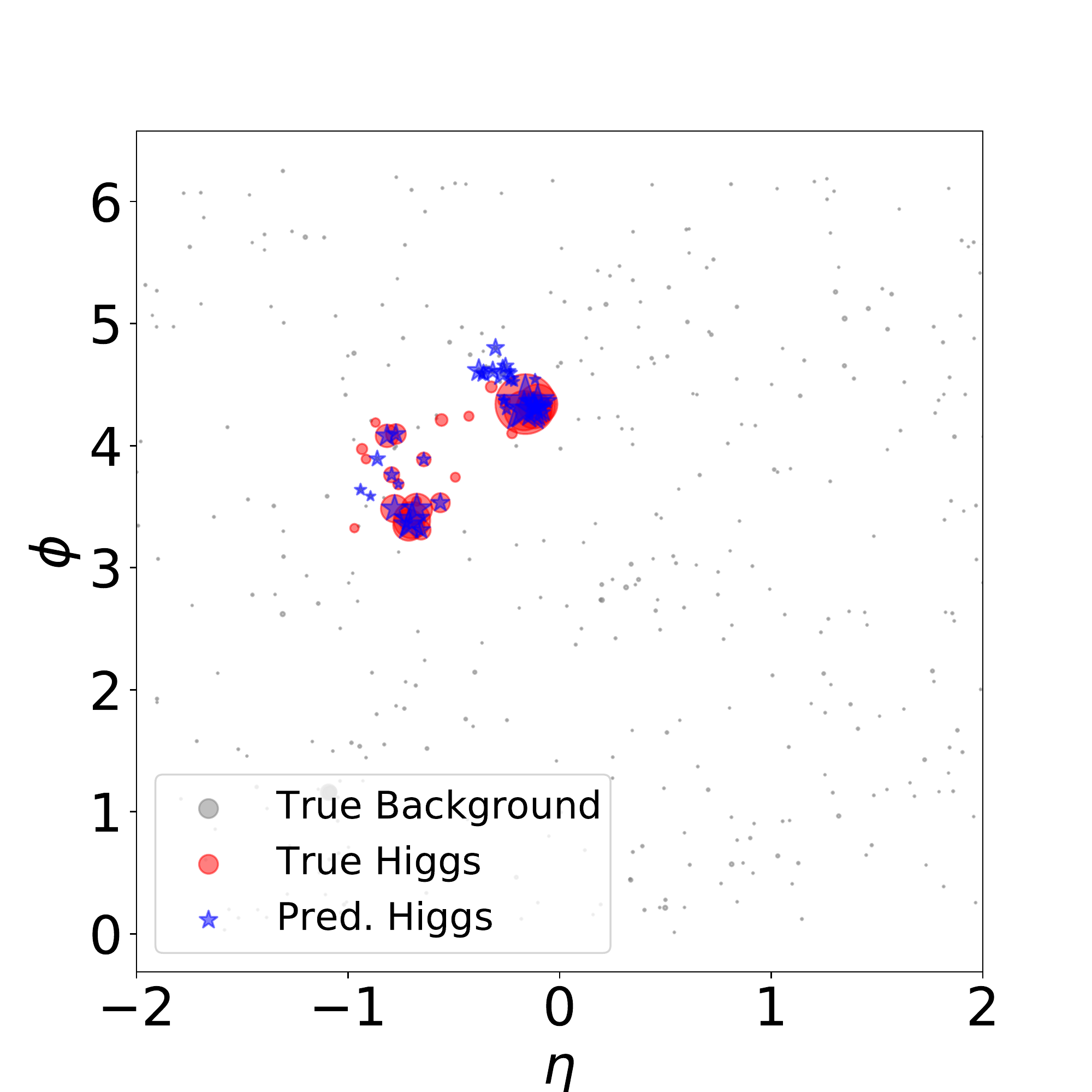}
     \includegraphics[width=0.42\textwidth]{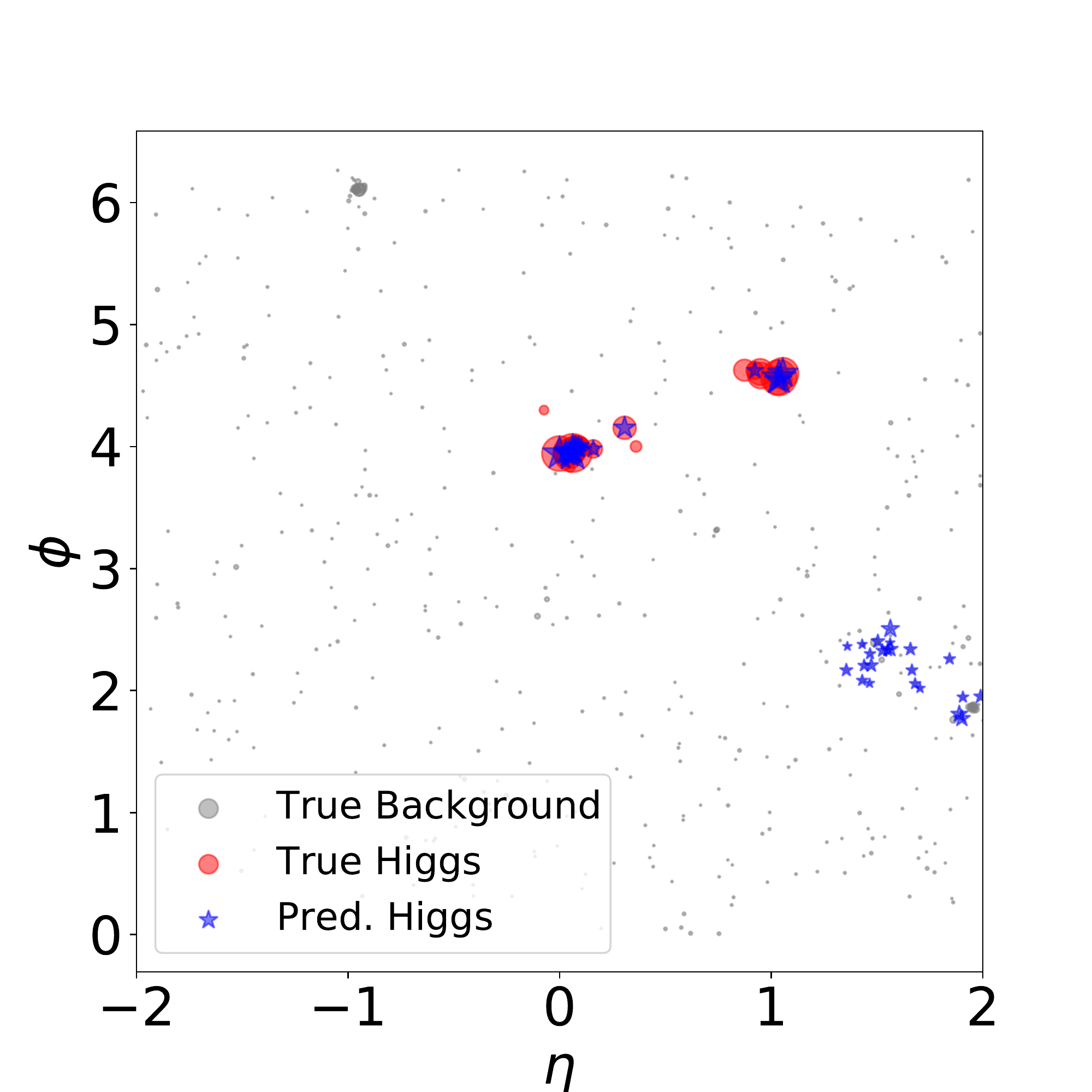}
    \caption{Each panel is showing a $H+$jets event in the $\eta - \phi$ plane, including the pileup contribution. The truth level Higgs constituents are marked by red disks and the predicted Higgs constituents are marked by blue stars. The size of disk and star is indicating the transverse momentum for each particle. For visibility, the sizes of the Higgs constituents (both truth level and GCN prediction) are magnified by a factor of 20.
 \label{fig:eventimage}}
  \end{center}
\end{figure}

The aim of our GCN is to mark the Higgs constituents out (with classification score greater than 0.5) from the particle list with heavy pileup contamination. 
The four-momentum of the reconstructed Higgs boson candidate is given by the sum of the four-momenta of the marked hadrons.  
In most cases, our GCN works efficiently, as shown in the left panel of Fig.~\ref{fig:eventimage}. 
While in other cases, particles with relatively large transverse momenta and far away from the Higgs in the $\eta - \phi$ plane will also be mis-assigned to the Higgs, as shown in the right panel of Fig.~\ref{fig:eventimage}.  So after the detection by the GCN, we further require that the Higgs constituents should lie within 
\be
\Delta R = \sqrt{(\eta_i - \eta_{H})^2 + (\phi_i - \phi_{H})^2} < 1.5 \label{eq:angsep}
\ee
of the reconstructed Higgs jet, otherwise the predicted Higgs constituents are unlabeled. 

To obtain an intuitive impression of the GCN performance, we compare it to a popular Higgs tagging method which is composed of Cambridge-Aachen (C/A) jet clustering~\cite{Dokshitzer:1997in}, mass-drop tagger~\cite{Butterworth:2008iy} and trimming~\cite{Krohn:2009th}, denoted by MDT in the following. 
In this method, the final state particles are clustered with the C/A jet algorithm with appropriate cone size parameter $R_0$ in order to capture most of the Higgs decay products.  They are referred to as fat jets.  
The mass-drop tagger uses two criteria to characterize the substructure of the fat jet: by undoing the jet clustering, there is at least one step which breaks the jet $j$ into two subjets $j_1$ and $j_2$ ($m_{j_1} > m_{j_2}$) such that the mass drop is significant ($m_{j_1} < 0.67 m_{j}$) and the splitting is not asymmetric $\frac{\min (p^2_{T}(j_1), p^2_T (j_2))}{ m^2_j} \Delta R^2_{j_1,j_2} > 0.09$. 
The trimming method selects the hard subjets inside a fat jet, in order to mitigate the pileup contamination. 
The cone size parameter of jet clustering and trimming parameters in the MDT analysis are optimized for each Higgs production process to achieve the highest reconstruction efficiency of Higgs jet within $125\pm 5$ GeV mass window.  

\begin{figure}
  \begin{center}
   \includegraphics[width=.40\textwidth]{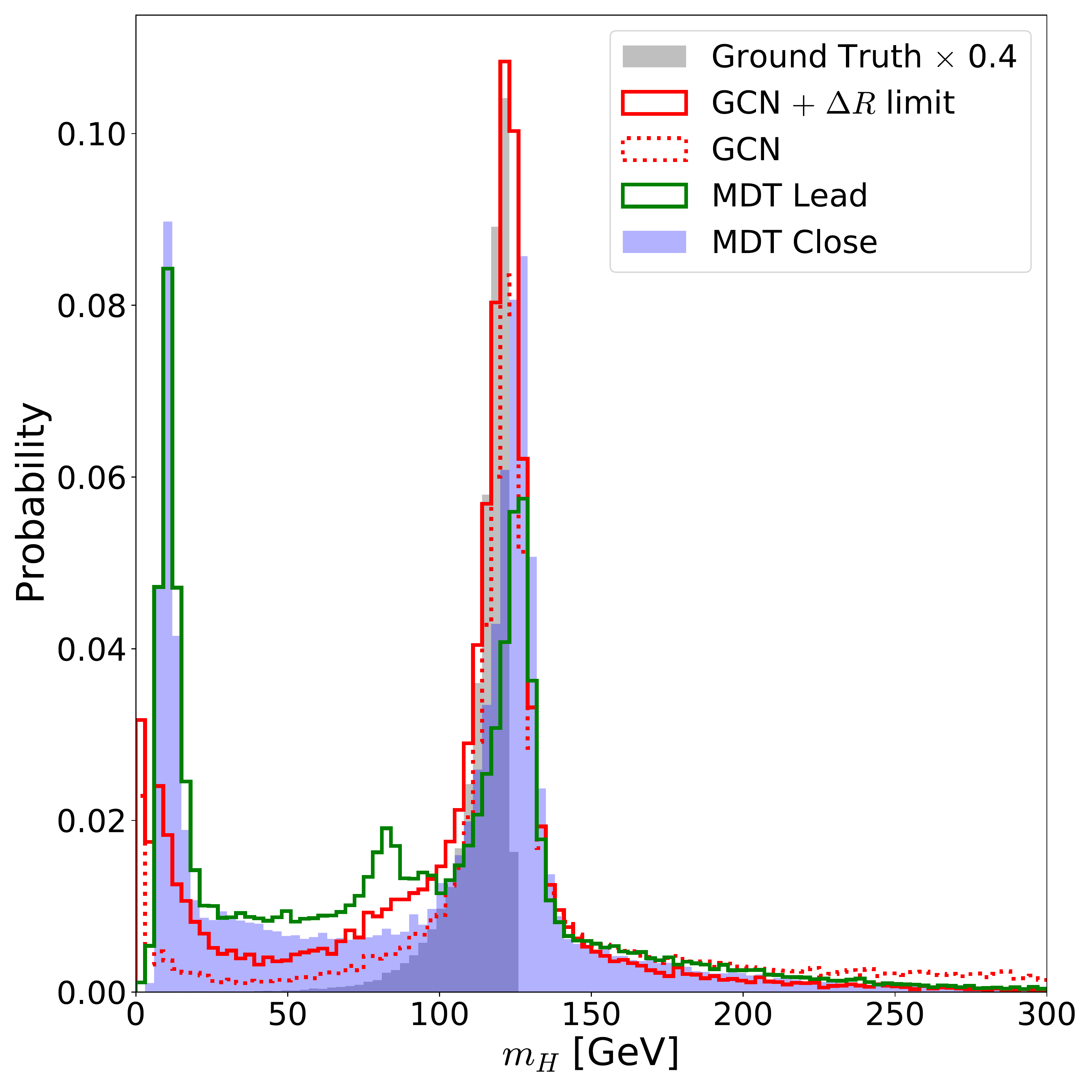}
     \includegraphics[width=.40\textwidth]{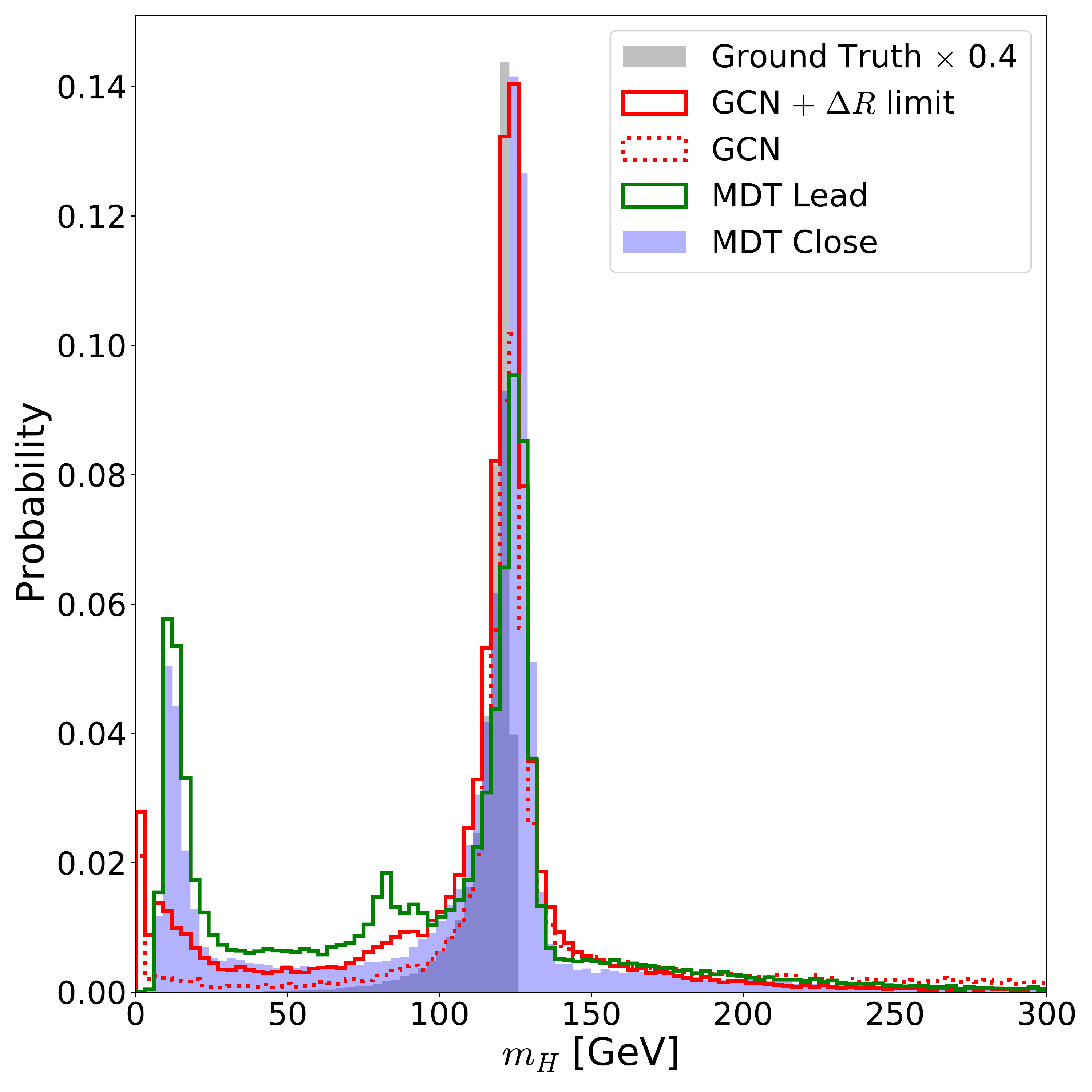}
    \caption{The invariant mass distribution of the selected Higgs constituents at ground truth level (the height of which is scaled by a factor of 0.4 for visibility), the reconstructed Higgs jet from the MDT method (by choosing the jet with highest $p_T$ and the jet closest to the true Higgs), and the GCN method (with and without the angular separation cut $\Delta R<1.5$). Methods applied to the event samples of  $H+$jets process with cut $p_T(H)>200$ GeV (left panel) and $p_T(H)>300$ GeV (right panel), respectively. \label{fig:im_distribution}}
  \end{center}
\end{figure}

In the Fig.~\ref{fig:im_distribution}, we plot the invariant mass distributions of the Higgs jets from the ground truth  selection, MDT analyses and GCN analyses, respectively. 
The ground truth selection corresponds to selecting the Higgs constituents which belong to the 1024 highest-$p_T$ particles for each event. The Higgs jet momentum is given by the vector sum of momenta of all selected Higgs constituents. In the figure, the height of histogram for the ground truth Higgs invariant mass is scaled by a factor of 0.4 for visibility. 
In the MDT analyses, $R_0=1.5$ (1.3) is chosen for fat jet clustering in event sample with Higgs $p_T>200$ (300) GeV. The fat jets are trimmed by  re-clustering the constituents into $R_\mathrm{sub}=0.20~(0.19)~k_t$-subjets and discarding those with $p_T^\mathrm{subjet}< f_{\rm cut} ~p_T^\mathrm{jet}$, where $f_{\rm cut}=0.05$.
Among those jets with substructure, the Higgs jet candidates are selected either as the leading $p_T$ trimmed fat jet (denoted by MDT Lead, which is oversimplified but realistic) or the one closest (in term of angular separation $\sqrt{\Delta \eta^2 + \Delta \phi^2}$) to the true Higgs boson (denoted by MDT Close, corresponds to the ideal case, which can be approached by using sophisticated Higgs tagging method). 
The performances of the GCN both with and without angular separation cut (Eq.~\ref{eq:angsep}) are shown. 
We can find that the GCN method with angular separation cut performs the best, in the sense that it has 
the largest number of events with reconstructed Higgs jet invariant mass close to 125 GeV.
In particular, when the Higgs decay products are well separated, the Higgs constituents can not be fully captured by the CA jet algorithm in the MDT Lead and MDT Close analyses. This leads to the peaks in the distributions of the reconstructed Higgs invariant mass at $\lesssim 10$ GeV in these two methods. 
The problem is less severe in the GCN methods. 
Comparing left and right panels of Fig.~\ref{fig:im_distribution}, because the higher transverse momentum of the Higgs lead to more remarkable Higgs jet features, the efficiency and the precision of the Higgs reconstructions are improved in all methods~\footnote{Although the parameters in the MDT method are adjusted accordingly for each event sample, the GCN is always the one trained on the event sample with $p_T>200$ GeV. }, and the fake peak at low invariant mass is greatly suppressed in the events sample with Higgs $p_T>300$ GeV. 

\begin{figure}
  \begin{center}
       \includegraphics[width=.40\textwidth]{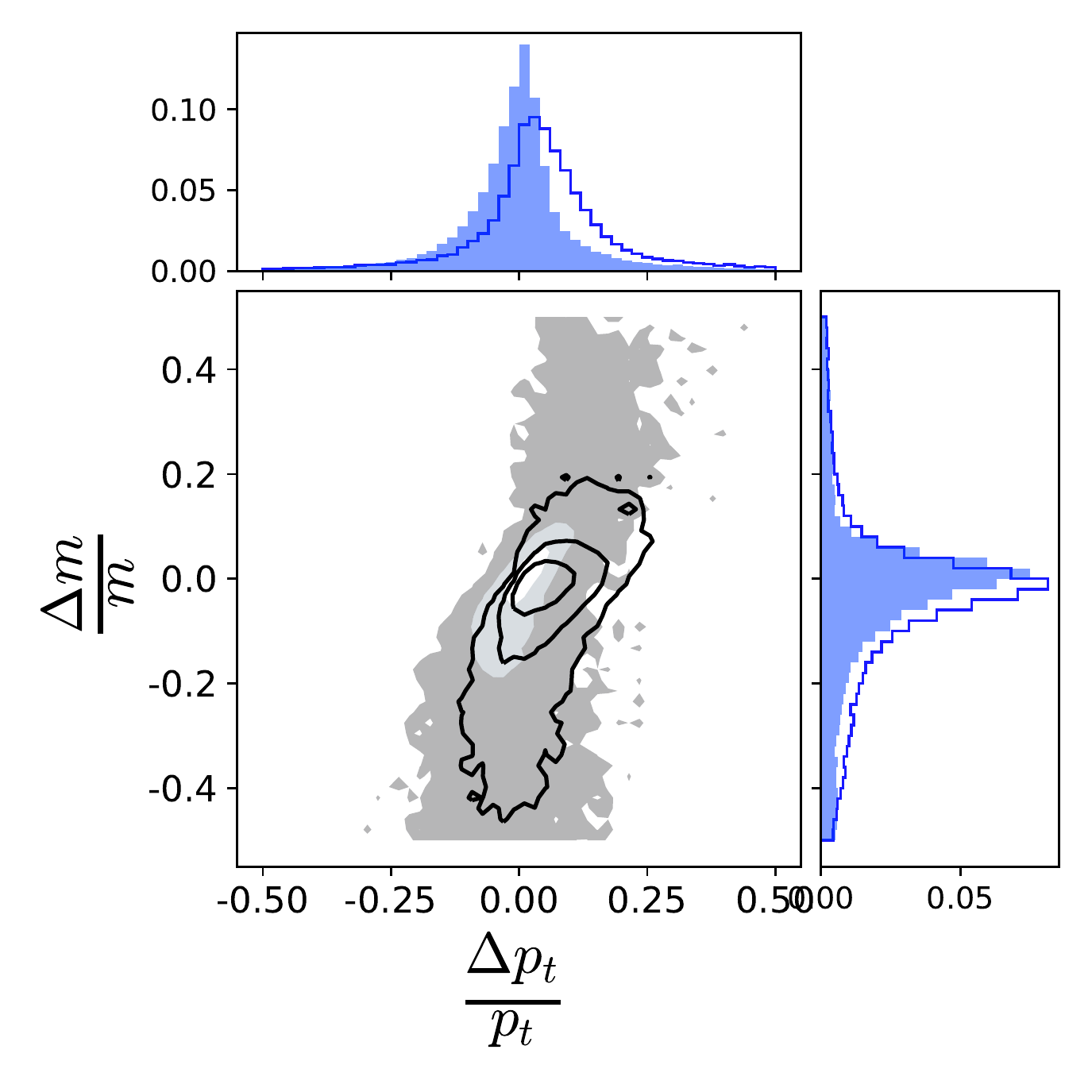}
     \includegraphics[width=.40\textwidth]{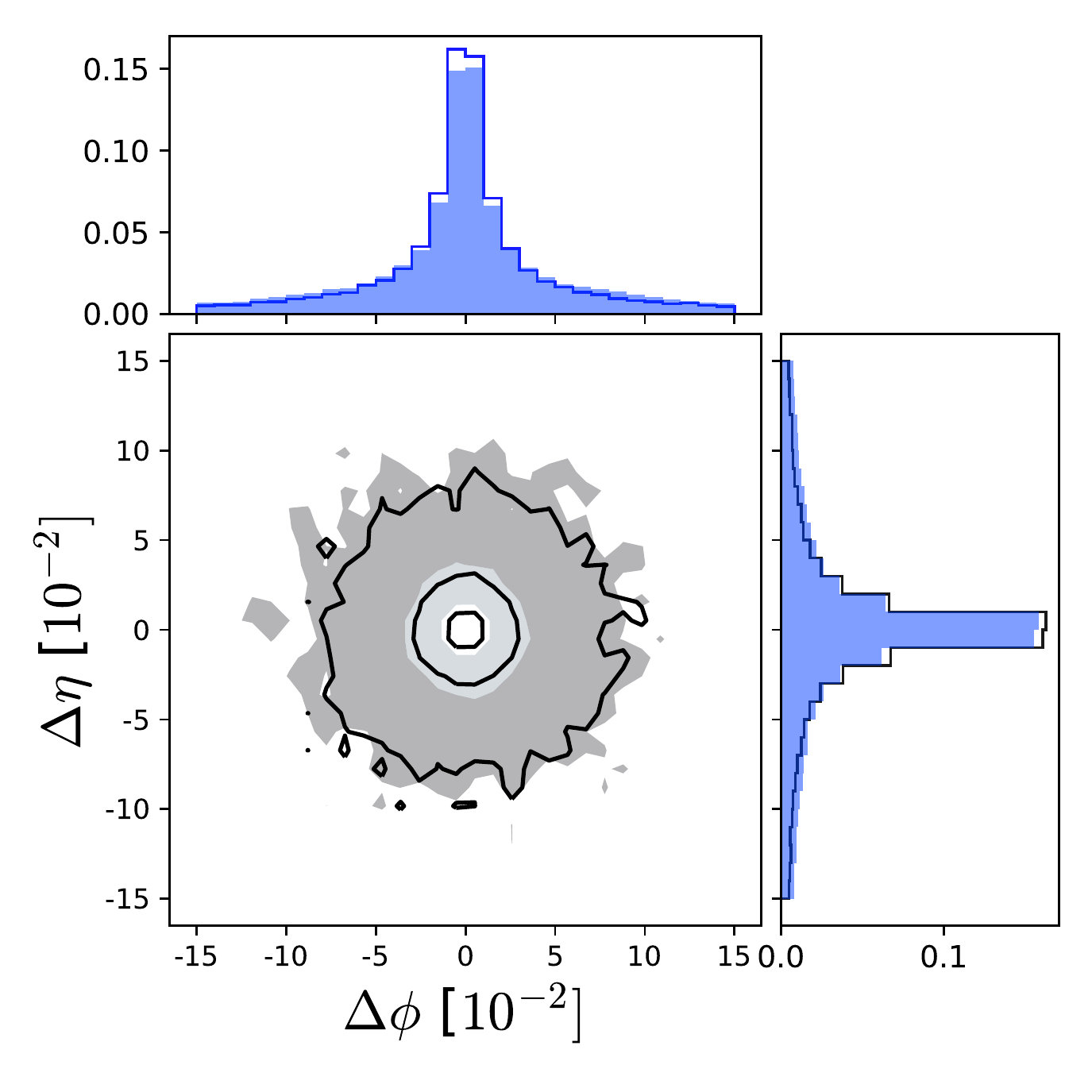}
     \includegraphics[width=.40\textwidth]{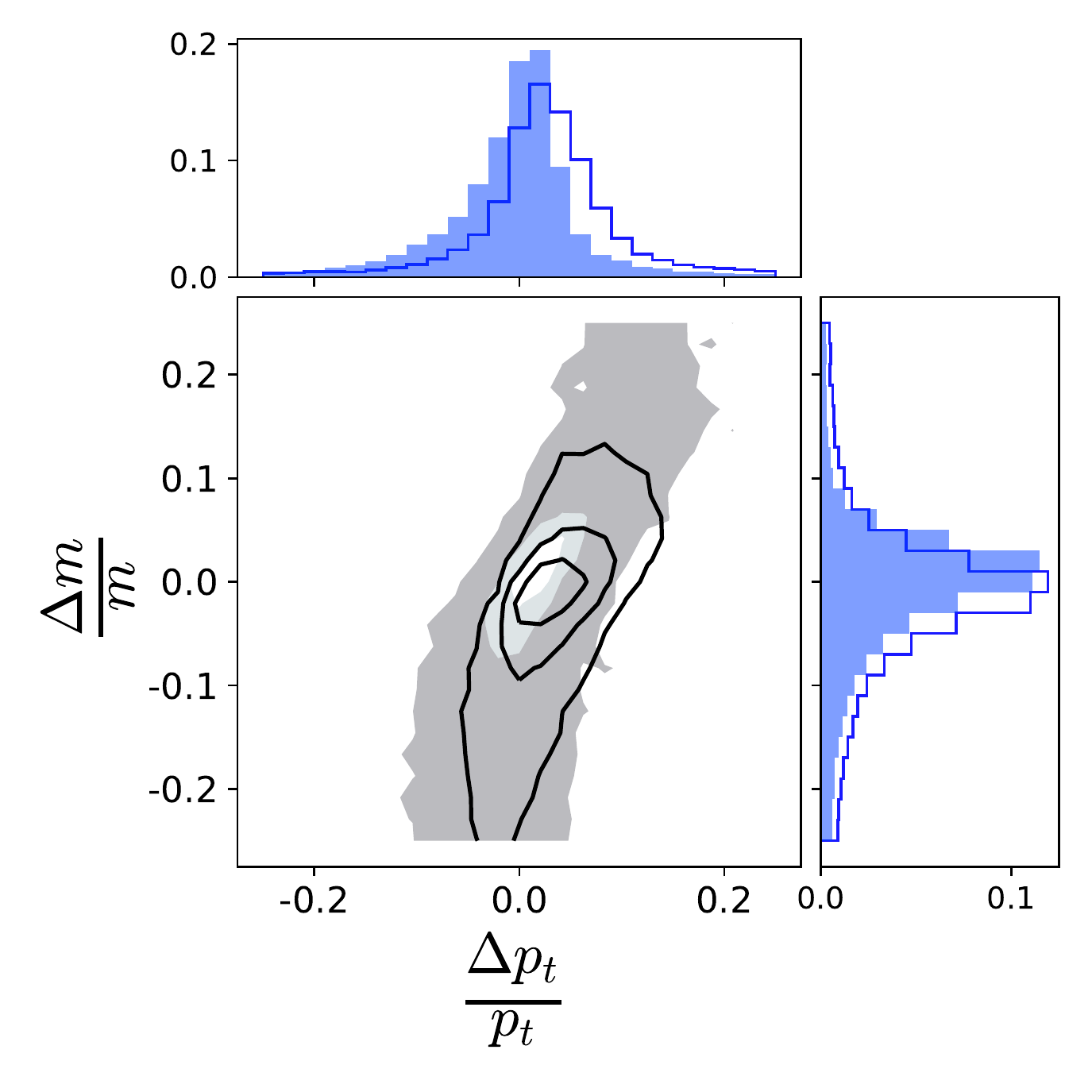}
     \includegraphics[width=.40\textwidth]{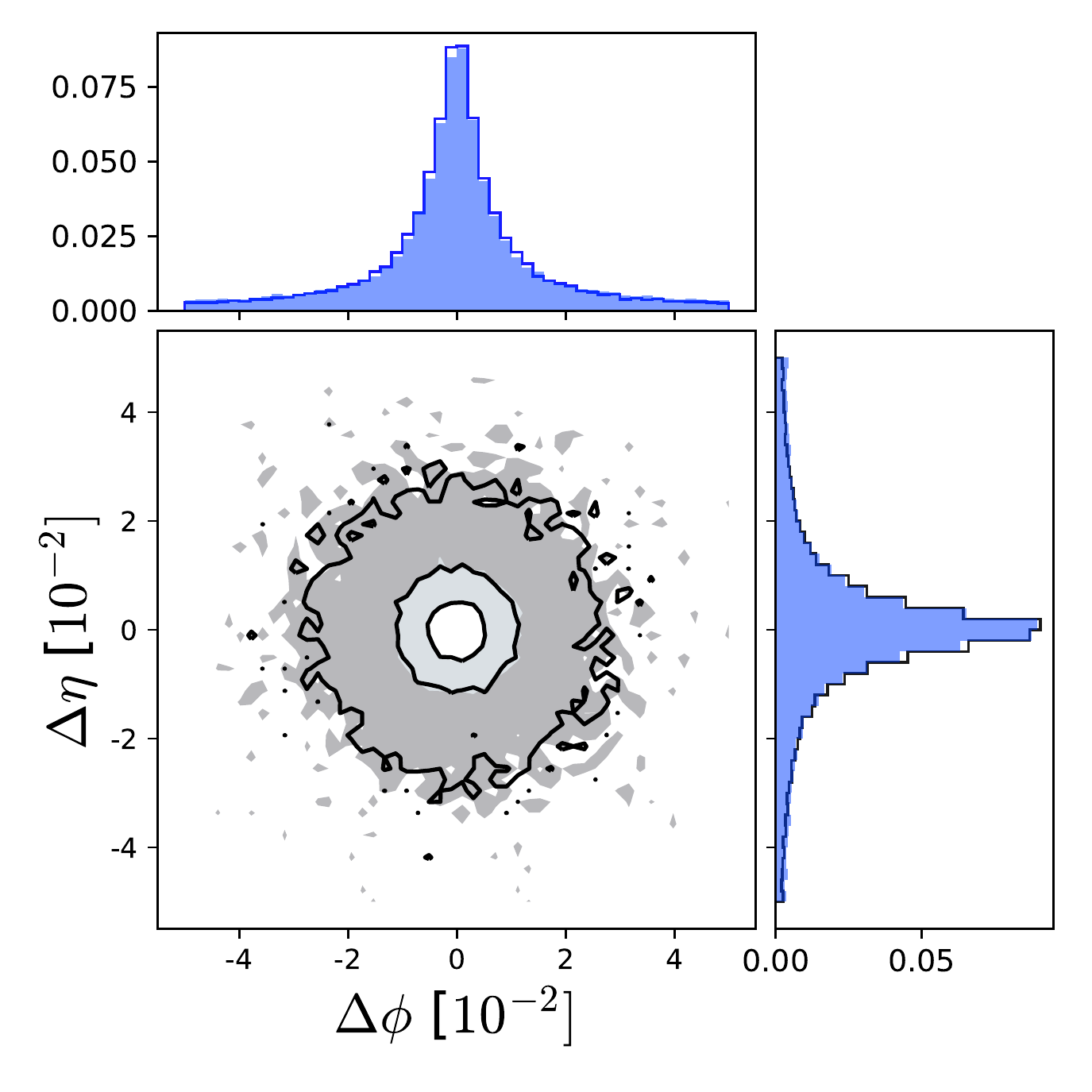}
    \caption{ $\frac{\Delta m}{m}- \frac{\Delta p_T}{p_T}$ and $\Delta \eta-\Delta \phi$ distributions for the reconstructed Higgs in the event samples of the $H+$jets process with cut $p_T(H)>200$ GeV (upper panels) and $p_T(H)>300$ GeV (lower panels). The grey (blue) regions indicate the distributions of reconstructed variables from the MDT Close method. And the black contours (blue lines) correspond to the distributions of reconstructed variables from the GCN method with $\Delta R$ cut.  \label{fig:deviations}}
  \end{center}
\end{figure}

The accuracies of the Higgs momentum reconstructions from the MDT Close method and the GCN method with $\Delta R$ limit are shown in Fig.~\ref{fig:deviations}, in terms of two-dimensional distributions on the $\frac{\Delta m}{m}- \frac{\Delta p_T}{p_T}$ plane and $\Delta \eta-\Delta \phi$ plane, where the $m=125$ GeV and $p_T$ is the Higgs boson transverse momentum at parton level; $\Delta m$, $\Delta p_T$, $\Delta \eta$ and $\Delta \phi$ are the deviations between the reconstructed Higgs momenta and the truth-level Higgs boson momentum. 
Two event samples of the $H+$jets process with $p_T(H)>200$ GeV and $p_T(H)>300$ GeV, are analyzed for illustration. 
The different shades of gray regions and black contours indicate 20\%, 40\% and 60\% of events, respectively. The closer they are to the center, the higher accuracy they stand for.
The GCN method outperforms the MDT method in kinematic variables reconstruction except the transverse momentum, where a systematic excess is prominent. 
Because the pileup effects are mitigated by trimming in the MDT method and no such pileup mitigation procedure is applied in the GCN method.  
We have used the vector sum of Higgs constituents momenta as the reconstructed Higgs momentum, the transverse momentum of the Higgs jet can be very sensitive to mis-assigned particles from pileup events, leading to an over estimated Higgs $p_T$.  
Using the GCN method to reconstruct the Higgs jet with $p_T>200$ GeV, more than half of the events can be reconstructed with $\frac{\Delta m}{m} \in [-0.4, 0.2]$,  $ \frac{\Delta p_T}{p_T} \in [-0.1,0.2]$, $\Delta \eta \in [-6,6] \times 10^{-2}$ and $\Delta \phi \in [-6,6] \times 10^{-2}$. The resolutions are even better for reconstructing Higgs jet with $p_T>300$ GeV.  

\subsection{Applications to different processes}
Even though the GCN is trained on the events of the $H+$jets process, it may serve as a general Higgs tagger for many other processes. For illustration, we perform the tests of the GCN on three processes different from the one used for training. Each process contains at least one Higgs jet in the final state. They are two Higgs plus jets process, Higgs plus a top quark pair process, and a SUSY process. The details of their simulations have been discussed in Sec.~\ref{sec:data}. 

\begin{figure}
  \begin{center}
     \includegraphics[width=.3\textwidth]{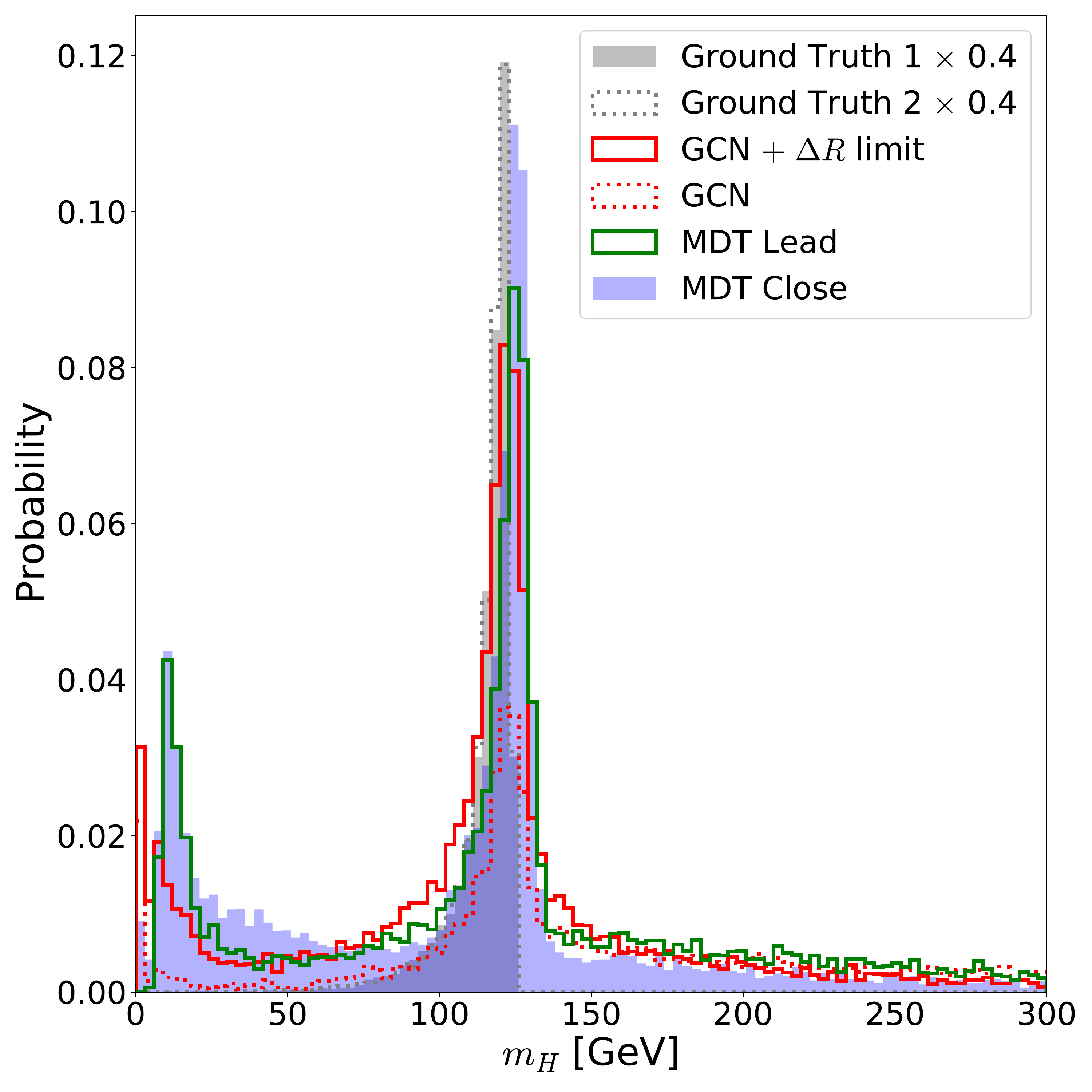}
     \includegraphics[width=.3\textwidth]{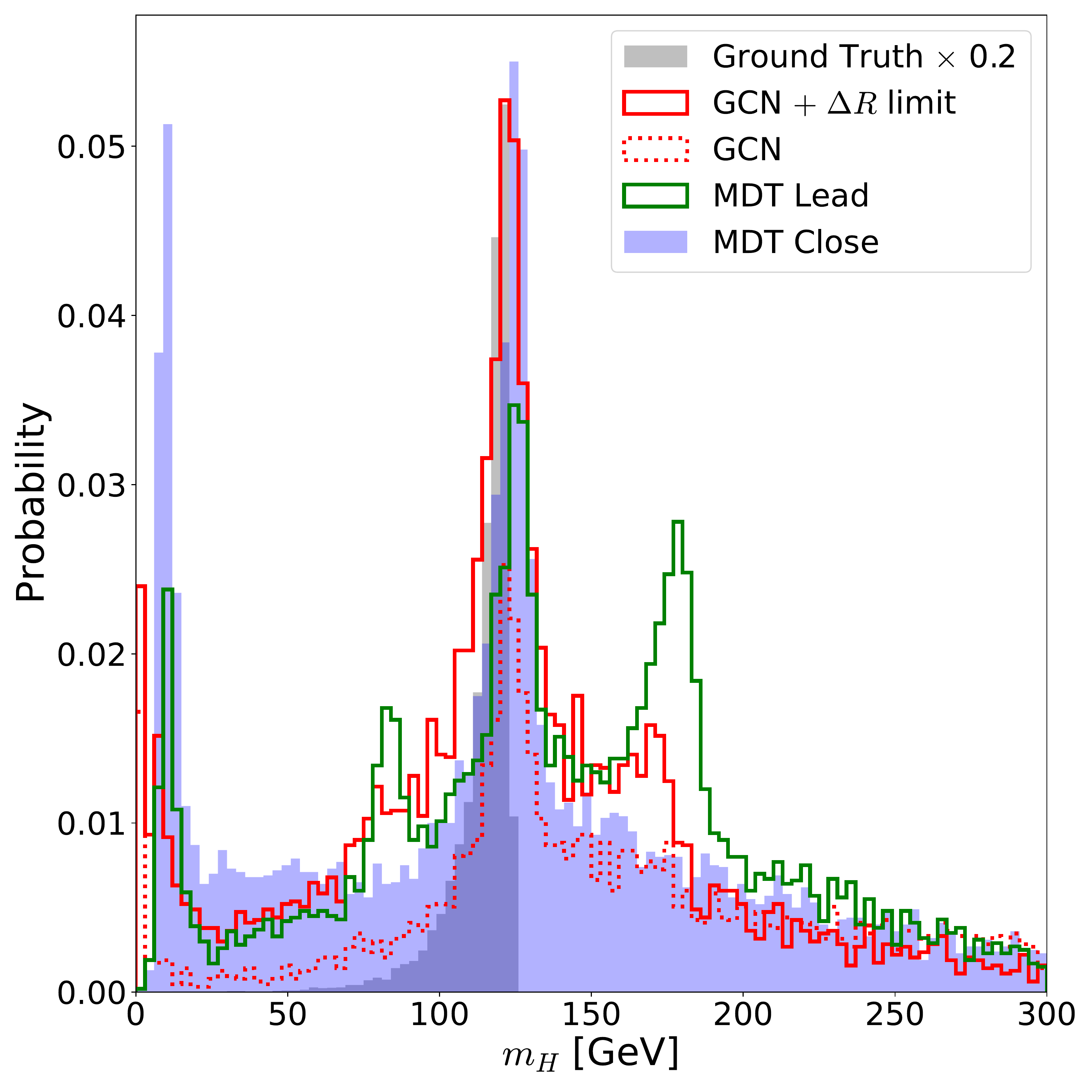}
     \includegraphics[width=.3\textwidth]{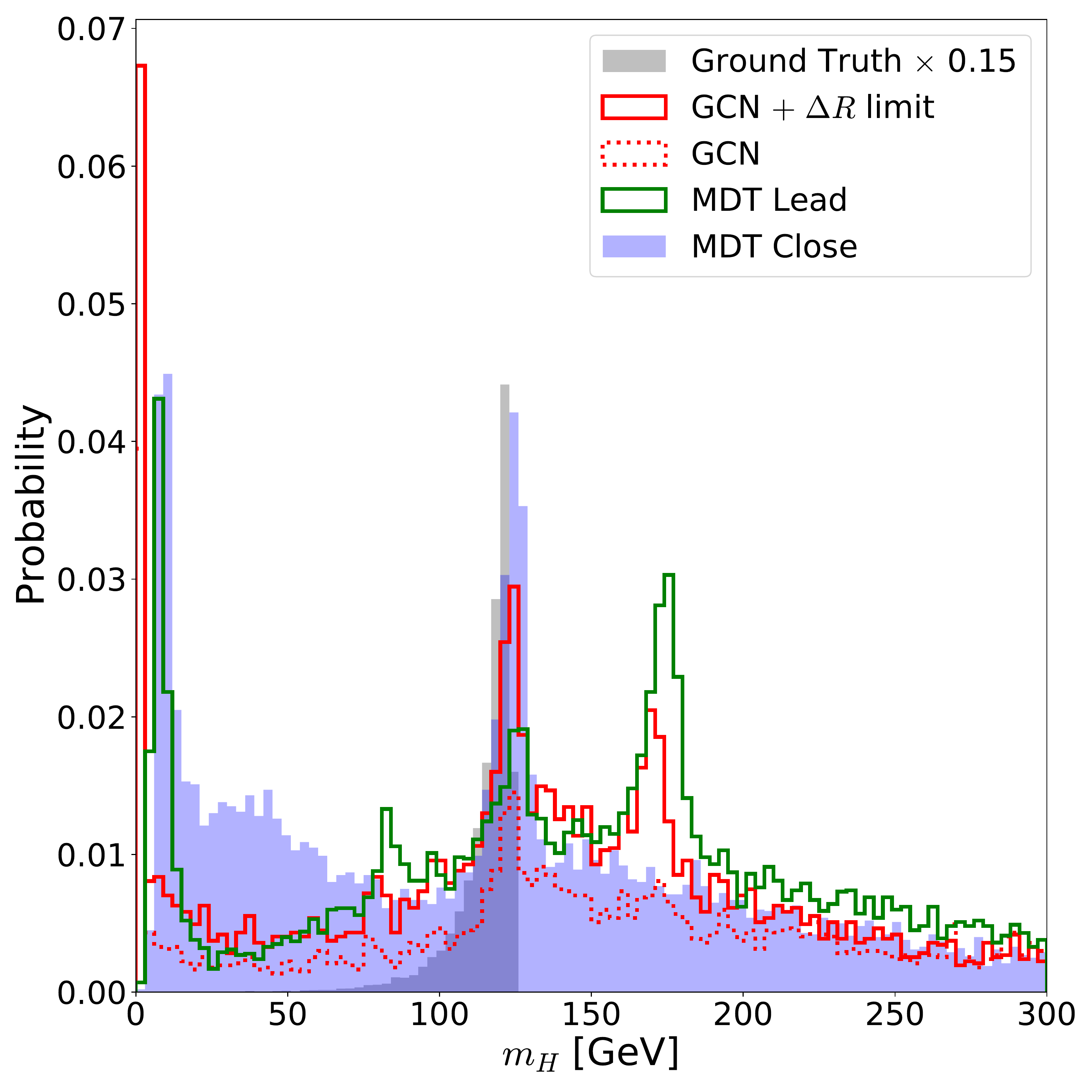}
    \caption{ Same as Fig.~\ref{fig:im_distribution}. Methods applied to the events samples of two Higgs plus jets process(left), Higgs plus a top quark pair process (middle), and a SUSY process (right).   \label{fig:3processes}}
  \end{center}
\end{figure}

In Fig.~\ref{fig:3processes}, we present the Higgs jet invariant mass distributions obtained by the MDT methods and the GCN methods for these three processes. As before, the ground truth distributions are superposed. 
For the two Higgs plus jets process, the Higgs jet candidate can be selected without ambiguity in GCN method and MDT Lead method. 
Unlike the image based method as proposed in Ref.~\cite{Li:2020grn} which can detect multiple Higgs jets in an event, the GCN is only capable of detecting one Higgs jet for each event. 
According to the left panel of Fig.~\ref{fig:3processes}, the GCN tends to find out the Higgs with more remarkable substructure in the Higgs pair production process. Thus, the peak of the invariant mass distribution at $\sim 125$ GeV is sharper than the one showing in the left panel of Fig.~\ref{fig:im_distribution}. 
While for the MDT Close method, the trimmed fat jet that is closest to either of the true Higgs bosons is chosen. 
The jet clustering and trimming parameters in the MDT analyses (same for MDT Lead and MDT Close) for different processes are optimized in the ranges of $R_0 \in [0.5,2.0]$, $R_{\text{sub}} \in [0.1,0.4]$ and $f_{\text{cut}} \in [0.02,0.1]$ with step sizes 0.1, 0.01 and 0.01, respectively.
The values with the highest reconstruction efficiency of Higgs jet within mass window $125\pm 5$ GeV~\footnote{Same values are obtained when optimizing the efficiency within the $125\pm 20$ GeV mass window.} are chosen, as shown in Tab.~\ref{tab:trim}.
Due to the stringent trimming parameters that we choose, part of the Higgs constituents are also removed in some events. It leads to the peaks of Higgs invariant mass distributions at around zero. The effect is more remarkable when there are larger number of energetic hadrons (which are not originated from the Higgs boson) flying around the Higgs jet.  

\begin{table}[h]
\centering
\begin{tabular}{|c|c|c|c|}
\hline
Process&$R_0$&$R_\text{sub}$&$f_\text{cut}$\\
\hline
$HHjjj$&1.4&0.19&0.05\\
$Ht\bar{t}$&1.4&0.19&0.05\\
$\tilde t_1\tilde t^*_1$&0.9&0.18&0.05\\
\hline
\end{tabular}
\caption{Jet clustering and trimming parameters in the MDT methods. The MDT Lead and MDT Close methods take the same values. \label{tab:trim}}
\end{table}



The performance of the GCN degrades in the second process due to the existence of energetic top quark decay products. 
While in the third process, one of the top quark is boosted, producing clusters of particles with high transverse momenta. It can be easily mis-assigned as Higgs constituents by the GCN (similar as in the right panel of Fig.~\ref{fig:eventimage}, GCN tends to assign high classification scores to the energetic particles within clusters), and they are too energetic to be thrown away by the angular separation cut $\Delta R<1.5$. As a result, the Higgs detection of the GCN method is not successful in the most of the cases for this process.  

In practice, if the target process has been set and for which the performance of our GCN is not satisfying, one can improve the performance by transfer learning:  initialize the model with our well trained weights; train the model by using events of the target process and with lower learning rate.

\subsection{Signal and Background Discrimination}

To study the background resisting capacity of the GCN, we carry out a comparison study on the performances of our GCN on the $H+$jets process with $p_T(H)>200$ (300) GeV and on the top quark pair production process.   
In simulation of the $t\bar{t}$ events, the parton level $t\bar{t}$ production is matched to parton shower up to two additional jets and each event is overlaid with average number of 50 pileup events.

The finite momentum resolutions are considered by smearing each component of a momentum with a Gaussian distribution. The standard derivatives (using the parameters of the ATLAS detector~\cite{Aharrouche:2006nf,Kulchitsky:2000gg}) for different particles are listed in Tab.~\ref{tab:smear}. 

\begin{table}[h!]
\centering 
\begin{tabular}{|c|c|c|c|c|c|c|} \hline
 &  $\sigma(\phi) = \sigma(\eta)$  & $\sigma(E)$ \\ \hline
Electron & 0.002 & $\sqrt{0.03^2 +(0.0013E)^2}$   \\ \hline
Muon &  0.001 &  $\sqrt{0.01^2 +(0.0001E)^2}$  \\ \hline
Tracker &  0.002 & $\sqrt{0.06^2 +(0.0013E)^2}$  \\ \hline
Photon & 0.02 &  $\sqrt{0.101^2E +(0.0017E)^2}$ \\ \hline
Neutral hadron & 0.1 &  $\sqrt{1.59^2 + 0.5205^2E +(0.0302E)^2}$ \\ \hline
\end{tabular}
\caption{The standard derivatives for the Gaussian smearing of azimuth, pseudorapidity and energy of different particles. \label{tab:smear}}
\end{table}

We adopt the Boosted Decision Tree (BDT) technique to obtain a combined discriminating powers of several variables. 
The variables are classified into two categories:
\begin{itemize}
\item Jet substructure variables for the Higgs jet candidate~\footnote{The variables in this class can be calculated for the Higgs jet candidate in both the MDT analysis and the GCN analysis. }. 
\item Variables reconstructed from the scores that the GCN assigns to Higgs constituents. 
\end{itemize} 

The first category includes the N-subjettiness variables~\cite{Thaler:2010tr} $\tau_{21}=\tau_2/\tau_1$ and $\tau_{32}=\tau_3/\tau_2$, with $\tau_N$ being defined as
\begin{align}
\tau_N \equiv \frac{\sum_k \min \{\Delta R_{1,k}, \Delta R_{2,k}, \cdots, \Delta R_{N,k}\}}{\sum_k p_{T,k} R_0}
\end{align}
where the summations run over all Higgs jet constituents, $R_0$ is the cone size parameter in the original jet clustering algorithm, $\Delta R_{I,k}$ denotes the angular distance ($\sqrt{\Delta \eta^2 + \Delta \phi^2}$) between the subjet $I$ and jet constituent $k$. There are also variables that characterize the jet profile: $f_{\rm cent}=E^{0.1}/E^{0.2}$ with $E^{0.1}/E^{0.2}$ is the summed energy of particles inside a cone of $R=0.1/0.2$ centered on the jet direction; the transverse momentum of the most energetic track inside a cone of radius $R = 0.2$ centered on the jet direction; $p_T$ weighted sum of the angular distances of all particles inside the jet. 

\begin{figure}[thb]
  \begin{center}
     \includegraphics[width=.32\textwidth]{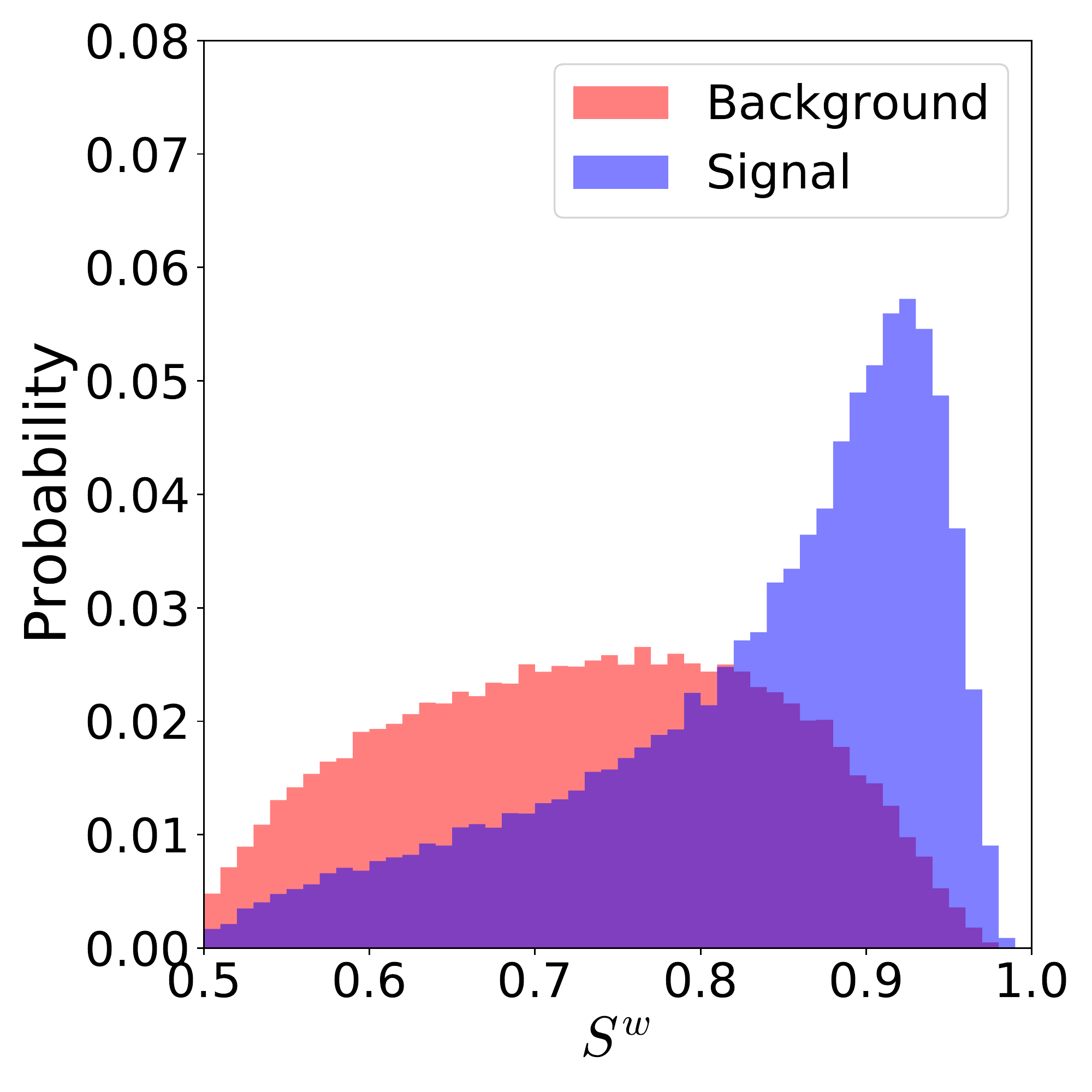}
     \includegraphics[width=.32\textwidth]{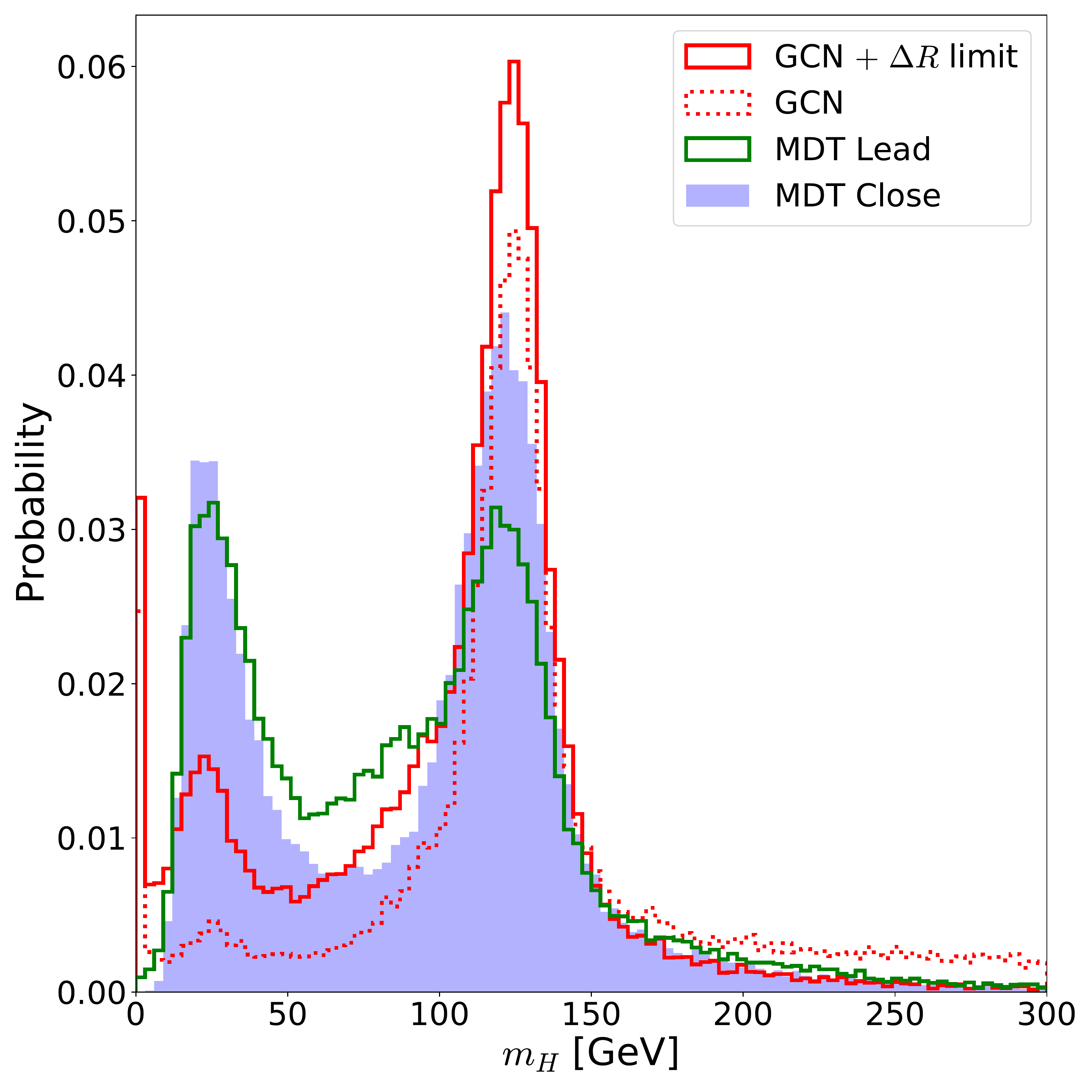}
     \includegraphics[width=.32\textwidth]{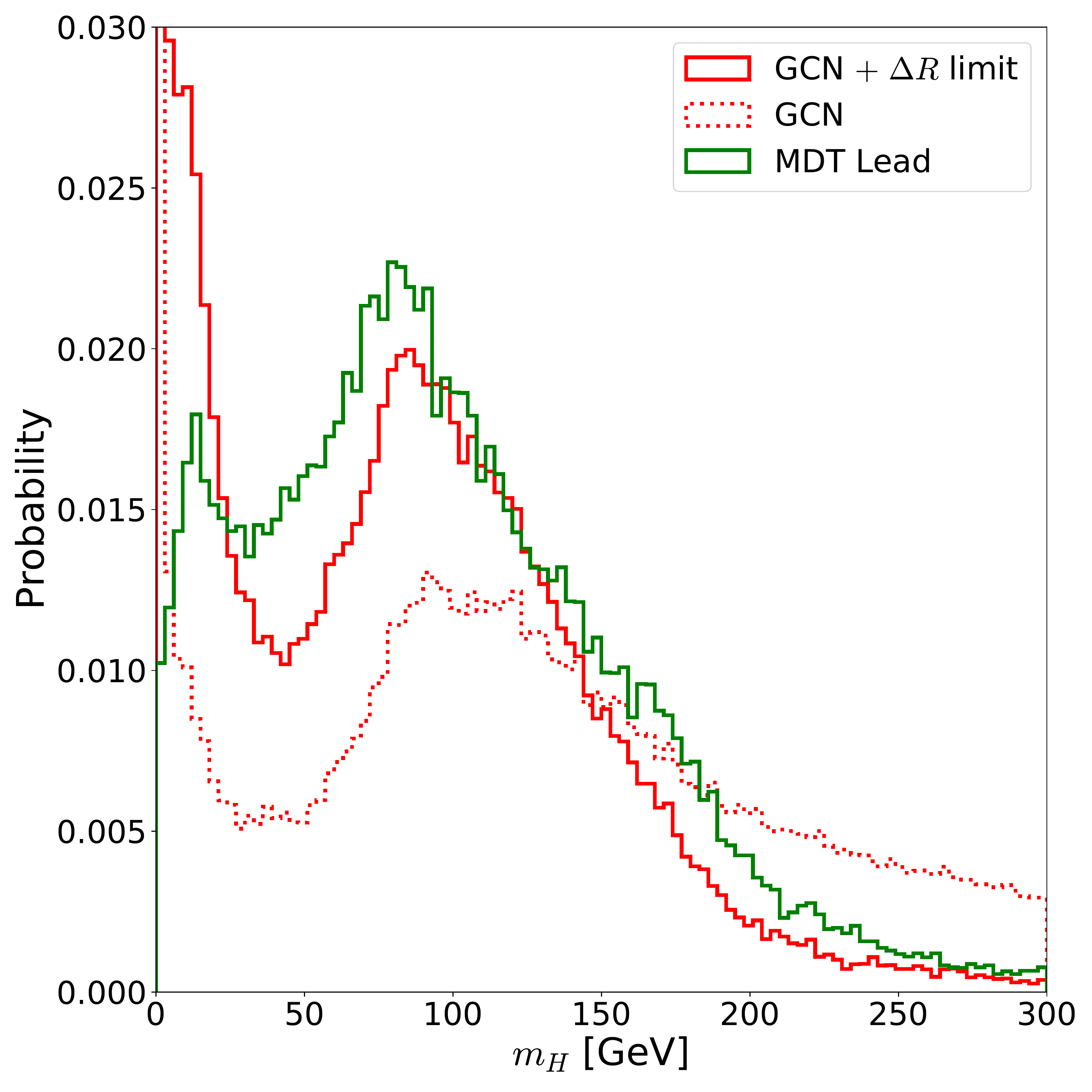}
    \caption{The distributions of the $p_T$ weighted score $S^w$ (left panel) and Higgs invariant mass for signal ($H$+jets with $p_T(H)>200$ GeV, middle panel) and $t\bar{t}$ background (right panel) processes.  The detector effects have been taken into account. 
    \label{fig:signal_background}}
  \end{center}
\end{figure}

The second category includes the average $\langle S \rangle$ and the standard deviation $\sigma_{S}$ of the scores that are assigned to the constituents of the Higgs jet by the GCN. Moreover, we define a $p_T$ weighted score $S^w$ for the Higgs jet candidate:
\be
S^w = \sum( s_i \cdot p_{T,i}) / \sum p_{T,i}
\ee
where $i$ runs over all Higgs constituents; $s_i$ and $p_{T,i}$ are the GCN score and the transverse momentum of the $i$th constituent. The distribution of the $S^w$ is presented in the left panel of Fig.~\ref{fig:signal_background}. 
Only events which contain GCN tagged Higgs jet are used. It includes 96\% of simulated signal events and 87\% of simulated background events.
The Higgs jet candidates in the events of the signal process tend to gain higher $S^w$  than that in the events of background process, which makes this variable useful for signal and background discrimination.

Beside the variables in those two categories, the transverse momentum $p_T(H)$ and invariant mass $m_H$ of the reconstructed Higgs are also powerful discriminating variables.
In the middle and right panels of Fig.~\ref{fig:signal_background}, we plot the distributions of reconstructed Higgs invariant mass for signal ($H$+jets with $p_T(H)>200$ GeV) and background ($t\bar{t}$) processes, taking into account the detector effects. 
For the signal process, compared with the left panel of Fig.~\ref{fig:im_distribution}, we can find that the detector effects broaden the peaks at 125 GeV in all methods. The GCN method still outperforms the MDT method after taking into account the detector effects. The distributions of the fake Higgs jet invariant mass for background are typically around $50$--$100$ GeV in MDT and GCN analyses, and the cut $\Delta R<1.5$ in the GCN method is found to be effective in suppressing the distribution in large $m_H$ region. 

\begin{figure}[thb]
  \begin{center}
     \includegraphics[width=.40\textwidth]{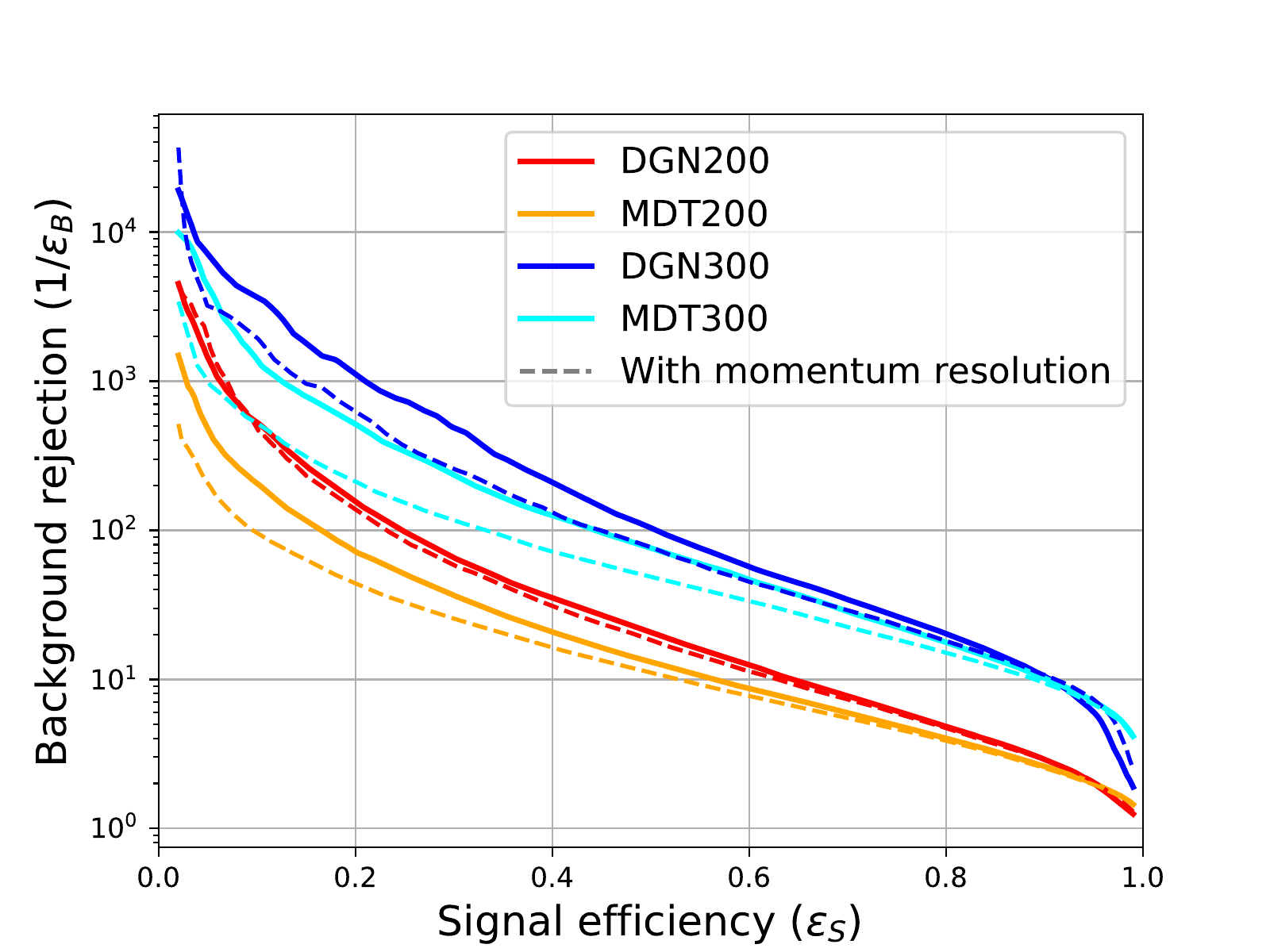}
     \includegraphics[width=.40\textwidth]{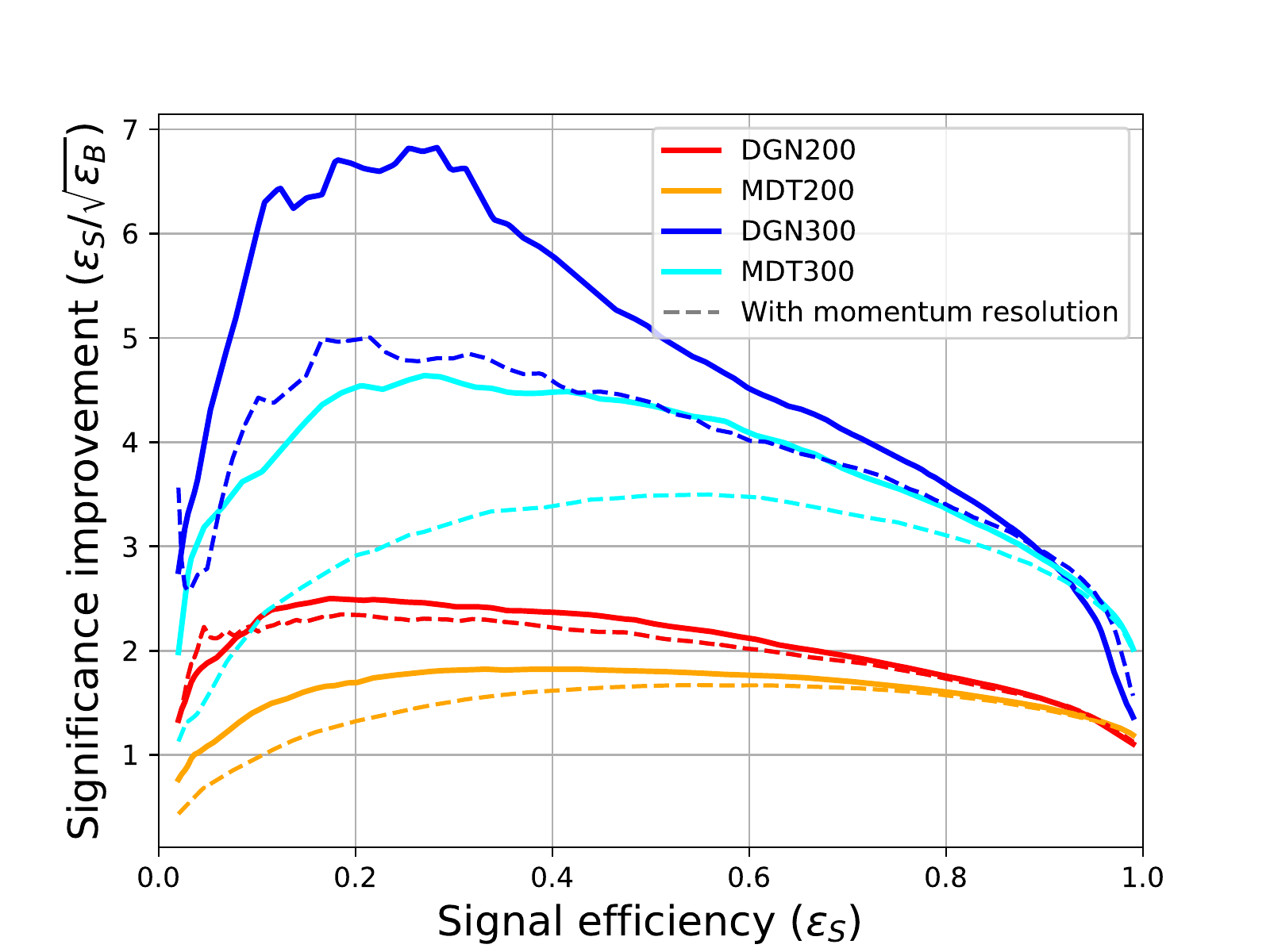}
    \caption{The receiver operating characteristic curves (left panel) and the significance improvement curves (right panel) of the GCN analyses and MDT analyses. Two different signal samples are considered: 
 $H+$jet with $p_T(H)>200$ GeV and $p_T(H)>300$ GeV. Results which take into account the finite momentum resolution are shown by the dashed lines. }
  \label{fig:roc}
  \end{center}
\end{figure}

In the following, the BDT analysis with input of variables in the first category as well as the $p_T(H)$ and the $m_H$ calculated on the MDT Lead tagged Higgs jet will be referred to as MDT analysis. And the BDT analysis with input of variables in both categories as well as the $p_T(H)$ and the $m_H$ calculated on the GCN tagged Higgs jet will be referred to as GCN analysis. 
The receiver operating characteristic (ROC)  curves for these two BDT analyses applying to two signal samples (with $p_T(H)>200$ GeV and $p_T(H)>300$ GeV, respectively) are plotted in the left panel of Fig.~\ref{fig:roc}.  
The solid and dashed lines correspond to the results with and without considering the finite momentum resolutions, respectively. 
We can find the GCN analysis has improved signal and background discriminating power than the MDT analysis for both signal samples. 
Moreover, the results obtained by the GCN analyses are less sensitive to the momentum resolution than the results of the MDT analyses. 
The significance improvement curves are presented in the right panel of Fig.~\ref{fig:roc}. It shows that the signal significances improve most when the signal selection efficiency $\epsilon_S \sim 0.2$. And the GCN analysis outperforms the MDT analysis by a factor of 1.5 on the signal significance for both signals.

\section{Conclusion and outlook}\label{sec:conclusion}

Representing each collider event as a point cloud, we adopt the GCN with focal loss to reconstruct the Higgs jet in the event. 
The $H+$jets with cut $p_T(H)>200$ GeV is taken as the benchmark process, on events of which the GCN is trained. 
We find the GCN analysis outperforms the traditional jet substructure analysis in both Higgs tagging efficiency and momentum reconstruction accuracy. And the GCN method is less sensitive to pileup contamination. 
In the GCN method without dedicated pileup mitigation, 60\% of $H+$jets events with cut $p_T>200$ (300) GeV can be reconstructed with $|\Delta \phi| \sim |\Delta \eta| < 6 \times 10^{-2} ~(2.5 \times 10^{-2})$, $\Delta m /m \in [-0.4, 0.2] ~([-0.25,0.12])$ and $\Delta p_T / p_T \in [-0.1,0.2] ~([-0.05,0.1])$.

The GCN which is trained on the events of the $H+$jets process is also capable of reconstructing the Higgs jet in events of other processes, even though its performance is degraded when there are boosted particles other than the Higgs boson in an event. 
Finally, we show that the features learned by the GCN are complementary to jet substructure variables in separating the events of the $H+$jets process and the $t\bar{t}$ process.

There are several limitations of the current GCN method: 
(1) The method can only reconstruct a single Higgs jet for each event (even in the event with multiple Higgs bosons);
(2) Although we try to propose a general Higgs jet detection method which does not depend on the Higgs production process, it turns out that the current method is not efficient in detecting the Higgs jet in processes where the Higgs is accompanied by other energetic particles. Since events of those new processes are not used in training, modifying the network and loss function can not help improving the performance. On the other hand, if the target process has been set, we can always improve the performance by transfer learning (probably with modified loss);
(3) The b-tagging is not considered in our study, which is a very important step towards the Higgs tagging. 
These subjects will be studied in future works.

\begin{acknowledgments}
This work was supported in part by the Fundamental Research Funds for the Central Universities, by the National Science Foundation of China under Grant No. 11905149, by Projects No. 11847612 and No. 11875062 supported by the National Natural Science Foundation of China, and by the Key Research Program of Frontier Science, Chinese Academy of Sciences.
\end{acknowledgments}

\bibliographystyle{jhep}
\bibliography{dgcnn_main}

\providecommand{\href}[2]{#2}\begingroup\raggedright\begin{thebibliography}{10}

\bibitem{Aad:2019mbh}
{\scshape ATLAS} collaboration, \emph{{Combined measurements of Higgs boson
  production and decay using up to $80$ fb$^{-1}$ of proton-proton collision
  data at $\sqrt{s}=$ 13 TeV collected with the ATLAS experiment}},
  \href{https://doi.org/10.1103/PhysRevD.101.012002}{\emph{Phys. Rev. D}
  {\bfseries 101} (2020) 012002}
  [\href{https://arxiv.org/abs/1909.02845}{{\ttfamily 1909.02845}}].

\bibitem{Sirunyan:2018koj}
{\scshape CMS} collaboration, \emph{{Combined measurements of Higgs boson
  couplings in proton-proton collisions at $\sqrt{s}=13\,\text {Te}\text {V}
  $}}, \href{https://doi.org/10.1140/epjc/s10052-019-6909-y}{\emph{Eur. Phys.
  J. C} {\bfseries 79} (2019) 421}
  [\href{https://arxiv.org/abs/1809.10733}{{\ttfamily 1809.10733}}].

\bibitem{Abdesselam:2010pt}
A.~Abdesselam et~al., \emph{{Boosted Objects: A Probe of Beyond the Standard
  Model Physics}},
  \href{https://doi.org/10.1140/epjc/s10052-011-1661-y}{\emph{Eur. Phys. J. C}
  {\bfseries 71} (2011) 1661}
  [\href{https://arxiv.org/abs/1012.5412}{{\ttfamily 1012.5412}}].

\bibitem{Altheimer:2012mn}
A.~Altheimer et~al., \emph{{Jet Substructure at the Tevatron and LHC: New
  results, new tools, new benchmarks}},
  \href{https://doi.org/10.1088/0954-3899/39/6/063001}{\emph{J. Phys. G}
  {\bfseries 39} (2012) 063001}
  [\href{https://arxiv.org/abs/1201.0008}{{\ttfamily 1201.0008}}].

\bibitem{Altheimer:2013yza}
A.~Altheimer et~al., \emph{{Boosted Objects and Jet Substructure at the LHC.
  Report of BOOST2012, held at IFIC Valencia, 23rd-27th of July 2012}},
  \href{https://doi.org/10.1140/epjc/s10052-014-2792-8}{\emph{Eur. Phys. J. C}
  {\bfseries 74} (2014) 2792}
  [\href{https://arxiv.org/abs/1311.2708}{{\ttfamily 1311.2708}}].

\bibitem{Adams:2015hiv}
D.~Adams et~al., \emph{{Towards an Understanding of the Correlations in Jet
  Substructure}},
  \href{https://doi.org/10.1140/epjc/s10052-015-3587-2}{\emph{Eur. Phys. J. C}
  {\bfseries 75} (2015) 409}
  [\href{https://arxiv.org/abs/1504.00679}{{\ttfamily 1504.00679}}].

\bibitem{Larkoski:2017jix}
A.J.~Larkoski, I.~Moult and B.~Nachman, \emph{{Jet Substructure at the Large
  Hadron Collider: A Review of Recent Advances in Theory and Machine
  Learning}}, \href{https://doi.org/10.1016/j.physrep.2019.11.001}{\emph{Phys.
  Rept.} {\bfseries 841} (2020) 1}
  [\href{https://arxiv.org/abs/1709.04464}{{\ttfamily 1709.04464}}].

\bibitem{Asquith:2018igt}
R.~Kogler et~al., \emph{{Jet Substructure at the Large Hadron Collider:
  Experimental Review}},
  \href{https://doi.org/10.1103/RevModPhys.91.045003}{\emph{Rev. Mod. Phys.}
  {\bfseries 91} (2019) 045003}
  [\href{https://arxiv.org/abs/1803.06991}{{\ttfamily 1803.06991}}].

\bibitem{Butterworth:2008iy}
J.M.~Butterworth, A.R.~Davison, M.~Rubin and G.P.~Salam, \emph{{Jet
  substructure as a new Higgs search channel at the LHC}},
  \href{https://doi.org/10.1103/PhysRevLett.100.242001}{\emph{Phys. Rev. Lett.}
  {\bfseries 100} (2008) 242001}
  [\href{https://arxiv.org/abs/0802.2470}{{\ttfamily 0802.2470}}].

\bibitem{Cacciari:2008gp}
M.~Cacciari, G.P.~Salam and G.~Soyez, \emph{{The anti-$k_t$ jet clustering
  algorithm}}, \href{https://doi.org/10.1088/1126-6708/2008/04/063}{\emph{JHEP}
  {\bfseries 04} (2008) 063} [\href{https://arxiv.org/abs/0802.1189}{{\ttfamily
  0802.1189}}].

\bibitem{Dokshitzer:1997in}
Y.L.~Dokshitzer, G.~Leder, S.~Moretti and B.~Webber, \emph{{Better jet
  clustering algorithms}},
  \href{https://doi.org/10.1088/1126-6708/1997/08/001}{\emph{JHEP} {\bfseries
  08} (1997) 001} [\href{https://arxiv.org/abs/hep-ph/9707323}{{\ttfamily
  hep-ph/9707323}}].

\bibitem{Krohn:2009th}
D.~Krohn, J.~Thaler and L.-T.~Wang, \emph{{Jet Trimming}},
  \href{https://doi.org/10.1007/JHEP02(2010)084}{\emph{JHEP} {\bfseries 02}
  (2010) 084} [\href{https://arxiv.org/abs/0912.1342}{{\ttfamily 0912.1342}}].

\bibitem{Ellis:2009su}
S.D.~Ellis, C.K.~Vermilion and J.R.~Walsh, \emph{{Techniques for improved heavy
  particle searches with jet substructure}},
  \href{https://doi.org/10.1103/PhysRevD.80.051501}{\emph{Phys. Rev. D}
  {\bfseries 80} (2009) 051501}
  [\href{https://arxiv.org/abs/0903.5081}{{\ttfamily 0903.5081}}].

\bibitem{Guest:2018yhq}
D.~Guest, K.~Cranmer and D.~Whiteson, \emph{{Deep Learning and its Application
  to LHC Physics}},
  \href{https://doi.org/10.1146/annurev-nucl-101917-021019}{\emph{Ann. Rev.
  Nucl. Part. Sci.} {\bfseries 68} (2018) 161}
  [\href{https://arxiv.org/abs/1806.11484}{{\ttfamily 1806.11484}}].

\bibitem{Albertsson:2018maf}
K.~Albertsson et~al., \emph{{Machine Learning in High Energy Physics Community
  White Paper}},
  \href{https://doi.org/10.1088/1742-6596/1085/2/022008}{\emph{J. Phys. Conf.
  Ser.} {\bfseries 1085} (2018) 022008}
  [\href{https://arxiv.org/abs/1807.02876}{{\ttfamily 1807.02876}}].

\bibitem{Radovic:2018dip}
A.~Radovic, M.~Williams, D.~Rousseau, M.~Kagan, D.~Bonacorsi, A.~Himmel et~al.,
  \emph{{Machine learning at the energy and intensity frontiers of particle
  physics}}, \href{https://doi.org/10.1038/s41586-018-0361-2}{\emph{Nature}
  {\bfseries 560} (2018) 41}.

\bibitem{Datta:2019ndh}
K.~Datta, A.~Larkoski and B.~Nachman, \emph{{Automating the Construction of Jet
  Observables with Machine Learning}},
  \href{https://doi.org/10.1103/PhysRevD.100.095016}{\emph{Phys. Rev. D}
  {\bfseries 100} (2019) 095016}
  [\href{https://arxiv.org/abs/1902.07180}{{\ttfamily 1902.07180}}].

\bibitem{Lin:2018cin}
J.~Lin, M.~Freytsis, I.~Moult and B.~Nachman, \emph{{Boosting $H\to b\bar b$
  with Machine Learning}},
  \href{https://doi.org/10.1007/JHEP10(2018)101}{\emph{JHEP} {\bfseries 10}
  (2018) 101} [\href{https://arxiv.org/abs/1807.10768}{{\ttfamily
  1807.10768}}].

\bibitem{Moreno:2019neq}
E.A.~Moreno, T.Q.~Nguyen, J.-R.~Vlimant, O.~Cerri, H.B.~Newman, A.~Periwal
  et~al., \emph{{Interaction networks for the identification of boosted $H
  \rightarrow b\overline{b}$ decays}},
  \href{https://doi.org/10.1103/PhysRevD.102.012010}{\emph{Phys. Rev. D}
  {\bfseries 102} (2020) 012010}
  [\href{https://arxiv.org/abs/1909.12285}{{\ttfamily 1909.12285}}].

\bibitem{Chakraborty:2019imr}
A.~Chakraborty, S.H.~Lim and M.M.~Nojiri, \emph{{Interpretable deep learning
  for two-prong jet classification with jet spectra}},
  \href{https://doi.org/10.1007/JHEP07(2019)135}{\emph{JHEP} {\bfseries 19}
  (2020) 135} [\href{https://arxiv.org/abs/1904.02092}{{\ttfamily
  1904.02092}}].

\bibitem{Sirunyan:2020lcu}
{\scshape CMS} collaboration, \emph{{Identification of heavy, energetic,
  hadronically decaying particles using machine-learning techniques}},
  \href{https://doi.org/10.1088/1748-0221/15/06/P06005}{\emph{JINST} {\bfseries
  15} (2020) P06005} [\href{https://arxiv.org/abs/2004.08262}{{\ttfamily
  2004.08262}}].

\bibitem{Li:2020grn}
J.~Li, T.~Li and F.-Z.~Xu, \emph{{Reconstructing boosted Higgs jets from event
  image segmentation}},  \href{https://arxiv.org/abs/2008.13529}{{\ttfamily
  2008.13529}}.

\bibitem{Shlomi:2020gdn}
J.~Shlomi, P.~Battaglia and J.-R.~Vlimant, \emph{{Graph Neural Networks in
  Particle Physics}},  \href{https://arxiv.org/abs/2007.13681}{{\ttfamily
  2007.13681}}.

\bibitem{Martinez:2018fwc}
J.~Arjona~Mart\'\i{}nez, O.~Cerri, M.~Pierini, M.~Spiropulu and J.-R.~Vlimant,
  \emph{{Pileup mitigation at the Large Hadron Collider with graph neural
  networks}}, \href{https://doi.org/10.1140/epjp/i2019-12710-3}{\emph{Eur.
  Phys. J. Plus} {\bfseries 134} (2019) 333}
  [\href{https://arxiv.org/abs/1810.07988}{{\ttfamily 1810.07988}}].

\bibitem{Farrell:2018cjr}
S.~Farrell et~al., \emph{{Novel deep learning methods for track
  reconstruction}},  in \emph{{4th International Workshop Connecting The Dots
  2018}}, 10, 2018 [\href{https://arxiv.org/abs/1810.06111}{{\ttfamily
  1810.06111}}].

\bibitem{Ju:2020xty}
X.~Ju et~al., \emph{{Graph Neural Networks for Particle Reconstruction in High
  Energy Physics detectors}},  in \emph{{33rd Annual Conference on Neural
  Information Processing Systems}}, 3, 2020
  [\href{https://arxiv.org/abs/2003.11603}{{\ttfamily 2003.11603}}].

\bibitem{Komiske:2018cqr}
P.T.~Komiske, E.M.~Metodiev and J.~Thaler, \emph{{Energy Flow Networks: Deep
  Sets for Particle Jets}},
  \href{https://doi.org/10.1007/JHEP01(2019)121}{\emph{JHEP} {\bfseries 01}
  (2019) 121} [\href{https://arxiv.org/abs/1810.05165}{{\ttfamily
  1810.05165}}].

\bibitem{Qu:2019gqs}
H.~Qu and L.~Gouskos, \emph{{ParticleNet: Jet Tagging via Particle Clouds}},
  \href{https://doi.org/10.1103/PhysRevD.101.056019}{\emph{Phys. Rev. D}
  {\bfseries 101} (2020) 056019}
  [\href{https://arxiv.org/abs/1902.08570}{{\ttfamily 1902.08570}}].

\bibitem{Moreno:2019bmu}
E.A.~Moreno, O.~Cerri, J.M.~Duarte, H.B.~Newman, T.Q.~Nguyen, A.~Periwal
  et~al., \emph{{JEDI-net: a jet identification algorithm based on interaction
  networks}}, \href{https://doi.org/10.1140/epjc/s10052-020-7608-4}{\emph{Eur.
  Phys. J. C} {\bfseries 80} (2020) 58}
  [\href{https://arxiv.org/abs/1908.05318}{{\ttfamily 1908.05318}}].

\bibitem{Mikuni:2020wpr}
V.~Mikuni and F.~Canelli, \emph{{ABCNet: An attention-based method for particle
  tagging}}, \href{https://doi.org/10.1140/epjp/s13360-020-00497-3}{\emph{Eur.
  Phys. J. Plus} {\bfseries 135} (2020) 463}
  [\href{https://arxiv.org/abs/2001.05311}{{\ttfamily 2001.05311}}].

\bibitem{Chakraborty:2020yfc}
A.~Chakraborty, S.H.~Lim, M.M.~Nojiri and M.~Takeuchi, \emph{{Neural
  Network-based Top Tagger with Two-Point Energy Correlations and Geometry of
  Soft Emissions}}, \href{https://doi.org/10.1007/JHEP07(2020)111}{\emph{JHEP}
  {\bfseries 20} (2020) 111}
  [\href{https://arxiv.org/abs/2003.11787}{{\ttfamily 2003.11787}}].

\bibitem{Bernreuther:2020vhm}
E.~Bernreuther, T.~Finke, F.~Kahlhoefer, M.~Kr\"amer and A.~M\"uck,
  \emph{{Casting a graph net to catch dark showers}},
  \href{https://arxiv.org/abs/2006.08639}{{\ttfamily 2006.08639}}.

\bibitem{Choma:2018zbe}
{\scshape IceCube} collaboration, \emph{{Graph Neural Networks for IceCube
  Signal Classification}},  \href{https://arxiv.org/abs/1809.06166}{{\ttfamily
  1809.06166}}.

\bibitem{Abdughani:2018wrw}
M.~Abdughani, J.~Ren, L.~Wu and J.M.~Yang, \emph{{Probing stop pair production
  at the LHC with graph neural networks}},
  \href{https://doi.org/10.1007/JHEP08(2019)055}{\emph{JHEP} {\bfseries 08}
  (2019) 055} [\href{https://arxiv.org/abs/1807.09088}{{\ttfamily
  1807.09088}}].

\bibitem{Ju:2020tbo}
X.~Ju and B.~Nachman, \emph{{Supervised Jet Clustering with Graph Neural
  Networks for Lorentz Boosted Bosons}},
  \href{https://arxiv.org/abs/2008.06064}{{\ttfamily 2008.06064}}.

\bibitem{Wang:2018nkf}
Y.~Wang, Y.~Sun, Z.~Liu, S.E.~Sarma, M.M.~Bronstein and J.M.~Solomon,
  \emph{{Dynamic Graph CNN for Learning on Point Clouds}},
  \href{https://arxiv.org/abs/1801.07829}{{\ttfamily 1801.07829}}.

\bibitem{2017arXiv170802002L}
T.-Y.~{Lin}, P.~{Goyal}, R.~{Girshick}, K.~{He} and P.~{Doll{\'a}r},
  \emph{{Focal Loss for Dense Object Detection}}, {\emph{arXiv e-prints} (2017)
  arXiv:1708.02002} [\href{https://arxiv.org/abs/1708.02002}{{\ttfamily
  1708.02002}}].

\bibitem{Alwall:2014hca}
J.~Alwall, R.~Frederix, S.~Frixione, V.~Hirschi, F.~Maltoni, O.~Mattelaer
  et~al., \emph{{The automated computation of tree-level and next-to-leading
  order differential cross sections, and their matching to parton shower
  simulations}}, \href{https://doi.org/10.1007/JHEP07(2014)079}{\emph{JHEP}
  {\bfseries 07} (2014) 079} [\href{https://arxiv.org/abs/1405.0301}{{\ttfamily
  1405.0301}}].

\bibitem{Sjostrand:2007gs}
T.~Sjostrand, S.~Mrenna and P.Z.~Skands, \emph{{A Brief Introduction to PYTHIA
  8.1}}, \href{https://doi.org/10.1016/j.cpc.2008.01.036}{\emph{Comput. Phys.
  Commun.} {\bfseries 178} (2008) 852}
  [\href{https://arxiv.org/abs/0710.3820}{{\ttfamily 0710.3820}}].

\bibitem{Skands:2014pea}
P.~Skands, S.~Carrazza and J.~Rojo, \emph{{Tuning PYTHIA 8.1: the Monash 2013
  Tune}}, \href{https://doi.org/10.1140/epjc/s10052-014-3024-y}{\emph{Eur.
  Phys. J. C} {\bfseries 74} (2014) 3024}
  [\href{https://arxiv.org/abs/1404.5630}{{\ttfamily 1404.5630}}].

\bibitem{ATLAS:2016puo}
{\scshape ATLAS} collaboration, \emph{{The Pythia 8 A3 tune description of
  ATLAS minimum bias and inelastic measurements incorporating the
  Donnachie-Landshoff diffractive model}}, .

\bibitem{Andersson:1983ia}
B.~Andersson, G.~Gustafson, G.~Ingelman and T.~Sjostrand, \emph{{Parton
  Fragmentation and String Dynamics}},
  \href{https://doi.org/10.1016/0370-1573(83)90080-7}{\emph{Phys. Rept.}
  {\bfseries 97} (1983) 31}.

\bibitem{Aad:2016mok}
{\scshape ATLAS} collaboration, \emph{{Charged-particle distributions in
  $\sqrt{s}$ = 13 TeV pp interactions measured with the ATLAS detector at the
  LHC}}, \href{https://doi.org/10.1016/j.physletb.2016.04.050}{\emph{Phys.
  Lett. B} {\bfseries 758} (2016) 67}
  [\href{https://arxiv.org/abs/1602.01633}{{\ttfamily 1602.01633}}].

\bibitem{Cornelis:2014ima}
{\scshape CMS} collaboration, \emph{{Quark-gluon Jet Discrimination At CMS}},
  in \emph{{2nd Large Hadron Collider Physics Conference}}, 9, 2014
  [\href{https://arxiv.org/abs/1409.3072}{{\ttfamily 1409.3072}}].

\bibitem{DBLP:journals/corr/JaderbergSZK15}
M.~Jaderberg, K.~Simonyan, A.~Zisserman and K.~Kavukcuoglu, \emph{Spatial
  transformer networks}, {\emph{CoRR} {\bfseries abs/1506.02025} (2015) }
  [\href{https://arxiv.org/abs/1506.02025}{{\ttfamily 1506.02025}}].

\bibitem{lin2018focal}
T.-Y.~Lin, P.~Goyal, R.~Girshick, K.~He and P.~Dollár, \emph{Focal loss for
  dense object detection},  2018.

\bibitem{10.5555/3026877.3026899}
M.~Abadi, P.~Barham, J.~Chen, Z.~Chen, A.~Davis, J.~Dean et~al.,
  \emph{Tensorflow: A system for large-scale machine learning},  in
  \emph{Proceedings of the 12th USENIX Conference on Operating Systems Design
  and Implementation}, OSDI'16, (USA), p.~265–283, USENIX Association, 2016.

\bibitem{Aharrouche:2006nf}
{\scshape ATLAS Electromagnetic Barrel Calorimeter} collaboration,
  \emph{{Energy linearity and resolution of the ATLAS electromagnetic barrel
  calorimeter in an electron test-beam}},
  \href{https://doi.org/10.1016/j.nima.2006.07.053}{\emph{Nucl. Instrum. Meth.
  A} {\bfseries 568} (2006) 601}
  [\href{https://arxiv.org/abs/physics/0608012}{{\ttfamily physics/0608012}}].

\bibitem{Kulchitsky:2000gg}
Y.A.~Kulchitsky, M.V.~Kuzmin, J.A.~Budagov, V.B.~Vinogradov and M.~Nessi,
  \emph{{Hadron energy reconstruction for the ATLAS barrel prototype combined
  calorimeter in the framework of the nonparametrical method}},
  \href{https://arxiv.org/abs/hep-ex/0004009}{{\ttfamily hep-ex/0004009}}.

\bibitem{Thaler:2010tr}
J.~Thaler and K.~Van~Tilburg, \emph{{Identifying Boosted Objects with
  N-subjettiness}}, \href{https://doi.org/10.1007/JHEP03(2011)015}{\emph{JHEP}
  {\bfseries 03} (2011) 015} [\href{https://arxiv.org/abs/1011.2268}{{\ttfamily
  1011.2268}}].

\end{thebibliography}\endgroup
\end{document}